%% file: main.tex
\documentclass[sigconf]{acmart}

\input{_header}

\begin{document}
\title{Keep the Lights On, Keep the Lengths in Check: Plug-In Adversarial Detection for Time-Series LLMs in Energy Forecasting}

\author{
Hua Ma$^\ast$,
Ruoxi Sun$^\ast$,  
Minhui Xue$^\ast$,
Xingliang Yuan$^\dagger$, 
Carsten Rudolph$^\ddagger$,
Surya Nepal$^\ast$,
Ling Liu$^\S$}
\affiliation{\institution{$^\ast$CSIRO's Data61, Australia}\country{}}
\affiliation{\institution{$^\dagger$The University of Melbourne, Australia}\country{}}
\affiliation{\institution{$^\ddagger$Monash University, Australia}\country{}}
\affiliation{\institution{$^\S$Georgia Institute of Technology, United States}\country{}}

\renewcommand \authors{Hua Ma, Ruoxi Sun, Minhui Xue, Xingliang Yuan, Carsten Rudolph, Surya Nepal, and Ling Liu}

\renewcommand{\shortauthors}{H. Ma, R. Sun, M. Xue, X. Yuan, C. Rudolph, S. Nepal, and L. Liu}

\begin{abstract}\label{sec:abstract} 
Accurate time-series forecasting is increasingly critical for planning and operations in low-carbon power systems. Emerging time-series large language models (TS-LLMs) now deliver this capability at scale, requiring no task-specific retraining, and are quickly becoming essential components within the Internet-of-Energy (IoE) ecosystem. However, their real-world deployment is complicated by a critical vulnerability: adversarial examples (AEs). Detecting these AEs is challenging because \textit{(i)} adversarial perturbations are optimized across the entire input sequence and exploit global temporal dependencies, which renders local detection methods ineffective, and \textit{(ii)} unlike traditional forecasting models with fixed input dimensions, TS-LLMs accept sequences of variable length, increasing variability that complicates detection. To address these challenges, we propose a \textit{plug-in} detection framework that capitalizes on the TS-LLM's own variable-length input capability. Our  method uses sampling-induced divergence as a detection signal. Given an input sequence, we generate multiple shortened variants and detect AEs by measuring the consistency of their forecasts: Benign sequences tend to produce stable predictions under sampling, whereas adversarial sequences show low forecast similarity, because perturbations optimized for a full-length sequence do not transfer reliably to shorter, differently-structured subsamples. We evaluate our approach on three representative TS-LLMs (TimeGPT, TimesFM, and TimeLLM) across three energy datasets: ETTh2 (Electricity Transformer Temperature), NI (Hourly Energy Consumption), and Consumption (Hourly Electricity Consumption and Production). Empirical results confirm strong and robust detection performance across both black-box and white-box attack scenarios, highlighting its practicality as a reliable safeguard for TS-LLM forecasting in real-world energy systems.
\end{abstract}



\maketitle

\section{Introduction}\label{sec_introduction} 

As global power systems advance toward deep decarbonization, accurate forecasting of demand, renewable generation, and market signals has become central to maintaining flexibility, reliability, and economic efficiency. AI-enabled forecasting already improves predictive accuracy, adaptability, and uncertainty quantification in areas like scheduling, reserve procurement, and market operations~\cite{shahverdi2025systematic,entezari2023artificial}. Similar progress in weather and climate modeling~\cite{wong2024ai} highlights the broader potential of learning-based approaches.

This evolution aligns with a broader structural shift from centralized grid operations toward Internet-of-Energy (IoE) systems, characterized by dense sensing, heterogeneous data streams, and increasingly automated decision processes~\cite{xue2023rai4ioe}. Modern IoE environments integrate high-resolution IoT meters, advanced metering infrastructure, and large fleets of distributed energy resources, producing massive volumes of time-series data describing load behavior, renewable output, asset conditions, and market behavior~\cite{dewangan2023load}. In such a distributed setting, machine learning has become essential for capturing nonlinear temporal dependencies and enabling real-time operational decisions beyond the reach of conventional tools. Practical adoption reflects this trend: smart-meter data supports load forecasting and anomaly detection~\cite{raza2015review}, probabilistic renewable forecasts guide reserve activation and frequency control~\cite{pinson2013wind,giebel2011state}, and data-driven analysis informs bidding optimization and manipulation detection~\cite{qiu2024robust}.

Large language models (LLMs) are increasingly finding a role within this IoE landscape. Their ability to process natural language allows for operator-assistant tools that interpret free-form utility logs and alarms~\cite{li2024large}, and SCADA records, providing readable summaries and explanations for control-room decisions~\cite{huang2024large}. In electricity markets, foundation models help surface abnormal bidding behavior and potential manipulation~\cite{alsharif2025energy}. At the distribution and residential scale, conversational LLM-based energy-management assistants translate user preferences into structured device-level actions~\cite{sanguinetti2024conversational}. LLM systems have also been explored for summarizing meteorological forecasts and interpreting wind and solar production patterns~\cite{liang2025energygpt}. Together, these developments indicate that LLMs are becoming an integral component of monitoring, control, and market workflows across distributed energy systems. This deep integration makes LLM-based forecasting models increasingly critical for real-world IoE operations.

Although LLMs were originally developed for natural-language tasks, growing evidence shows that they generalize effectively to time-series data~\cite{jin2023time,tan2024language,liu2025timecma,liu2025calf,gruver2023large}, delivering measurable gains in weather forecasting and energy-demand prediction~\cite{jiang2024empowering,jin2023large}. This capability, combined with their increasing use in IoE operations, has positioned LLMs as emerging general-purpose forecasters in real-world energy systems. Current adaptation approaches either repurpose pretrained language models through prompt engineering or train LLMs directly on large-scale time-series corpora with temporally aligned tokenization~\cite{liang2024foundation}. Commercial TS-LLMs such as TimeGPT~\cite{garza2023timegpt}, TimesFM~\cite{das2024decoder}, Chronos~\cite{ansari2024chronos}, and Moirai~\cite{woo2024unified} follow this paradigm, providing strong zero-shot performance and enabling domain users across finance, healthcare, and energy management to obtain accurate forecasts through general-purpose APIs. Their accelerating adoption within IoE workflows underscores the importance of ensuring that TS-LLM-based forecasting remains trustworthy and robust in operational settings.

Despite their rapid adoption, TS-LLM-based forecasting systems introduce significant challenges and new security risks. Unlike traditional models with fixed architectures and well-understood failure modes, TS-LLMs operate as large, opaque sequence learners whose outputs can be highly sensitive to input perturbations, data drift, and prompt variations~\cite{jin2023time,gruver2023large}. Recent studies show that they inherit the adversarial vulnerabilities of deep learning systems~\cite{jiang2024empowering} and can exhibit unstable behavior under subtle input changes, distribution shifts, or numerical encoding biases~\cite{tan2024language}.
These vulnerabilities become more pronounced in IoE settings, where data can be noisy, inconsistent across devices, and subject to real-time operational constraints. The difficulty is further heightened for commercial TS-LLMs, which are accessible only through fixed APIs and cannot be retrained or modified. Consequently, practical defenses must operate as modular, architecture-agnostic components that integrate seamlessly into existing TS-LLM forecasting pipelines without requiring access to model internals.

Motivated by these considerations, we examine the following key research question:

\begin{mdframed}[backgroundcolor=black!10,rightline=false,leftline=false,topline=false,bottomline=false,roundcorner=2mm]
\textit{Is there an effective adversarial example defense that can be used as a plug-in for emerging TS-LLMs without requiring any prior knowledge of adversarial attacks?}
\end{mdframed}

To resolve this, we exploit a distinctive property of TS-LLMs that is absent in conventional deep learning models: their ability to process inputs of flexible length. Building on this characteristic, we design a detection mechanism for identifying adversarial time-series inputs. Our second insight concerns the nature of adversarial perturbations, which are optimized on the full input sequence. When an adversarial input $\mathbf{x}$ is randomly partitioned into equal-length subsamples $\mathbf{x}_{\rm sub}$ and processed independently, the resulting forecasts diverge because perturbations tailored to the full sequence do not transfer reliably to the subsamples. Benign inputs lack such engineered modifications and therefore produce consistent forecasts across subsamples. This contrast forms the basis of our detection approach.

We summarize our main contributions as follows:
\begin{itemize}[leftmargin=*]
\item To the best of our knowledge, this is the first work to develop countermeasures against adversarial example attacks on TS-LLMs~\cite{liu2024adversarial}. We leverage the unique property of flexible input length in TS-LLMs to devise a length independence-based detection module (\name) that can be used as a seamlessly plug-in without modifying the existing TS-LLM systems.

\item \name supports various sampling and thresholding strategies to distinguish adversarial from benign inputs. Notably, the threshold is computed solely from benign data, which enables generalizable detection and increases robustness against adaptive attacks due to the inherent randomness of the sampling process.

\item We validate the effectiveness of \name through extensive experiments on 3 energy benchmark datasets using TimeGPT, TimeLLM, and TimesFM. Our evaluation considers both practical black-box attacks and white-box attacks. Across all settings, the underlying TS-LLMs remain intact, demonstrating the modular design and practicality of our defense.

\item Our evaluation focuses on black-box adversarial attacks to reflect realistic IoE conditions, and the method also demonstrates strong performance under white-box settings. We further show that adaptive thresholding can further improve detection accuracy.

\end{itemize}

\section{Related Work}\label{sec_related_work}

In this section, we introduce prior work on TS LLMs and on adversarial attacks and defenses relevant to time series forecasting.

\subsection{Time Series LLMs}\label{sec_time_series_llms}

Existing approaches for enabling LLMs to perform time-series forecasting follow two main directions. The first adapts pretrained LLMs to numerical sequences. OFA~\cite{zhou2023one} freezes most transformer layers and fine-tunes lightweight components for multiple forecasting tasks. PromptCast~\cite{xue2023promptcast} formulates forecasting as prompt-based generation, while TEMPO~\cite{cao2023tempo} combines trend–seasonality–residual decomposition with soft prompt pooling. TEST~\cite{sun2023test} leverages contrastive learning to align time-series tokens with text prototypes and LLM4TS~\cite{chang2023llm4ts} aligns LLMs with temporal structure prior to parameter-efficient adaptation. These approaches, however, typically require nontrivial fine-tuning and depend on tokenization schemes not naturally suited to continuous signals. To reduce adaptation overhead, LLMTime~\cite{gruver2023large} encodes values as digit strings to avoid training, and TimeLLM~\cite{jin2023time} introduces a lightweight reprogramming module that maps time series into and out of LLM-compatible representations.

The second direction trains foundation models directly on large-scale time-series corpora. TimeGPT~\cite{garza2023timegpt} was the first commercial model to demonstrate zero-shot forecasting, and Lag-LLaMA~\cite{rasul2023lag} extended this idea to probabilistic univariate prediction. More recent foundation models, including TimesFM~\cite{das2024decoder}, Chronos~\cite{ansari2024chronos}, and Moirai~\cite{woo2024unified}, further improve long-horizon forecasting and support flexible input and output lengths. By learning temporal structure from scratch, these TS-LLMs achieve strong zero-shot performance across multiple domains.

\subsection{Adversarial Example Attacks and Defenses}\label{sec_adv_and_def}

Adversarial examples were first identified in image models~\cite{szegedy2013intriguing}, motivating extensive work on both white-box and black-box attacks. Gradient-based methods such as FGSM and BIM~\cite{kurakin2018adversarial} dominate white-box settings, while black-box attacks typically rely on substitute models or evolutionary strategies~\cite{papernot2016transferability,su2019one}. GAN-based techniques have also been explored for perturbation generation~\cite{sarkar2017upset}. Although adversarial research in time series is comparatively limited, several studies have shown vulnerabilities in classification models: distance-based classifiers can be misled by crafted perturbations~\cite{oregi2018adversarial}, and FGSM or BIM can markedly degrade deep time-series classifiers~\cite{fawaz2019adversarial}. Subsequent work examined targeted and universal attacks~\cite{rathore2020untargeted} and developed substitute-model-based strategies~\cite{karim2020adversarial}. Attacks on forecasting models remain less explored, with examples including perturbations to autoregressive predictors~\cite{dang2020adversarial}, gradient-based perturbations guided by importance metrics~\cite{wu2022small}, and temporally targeted attacks~\cite{govindarajulu2023targeted}. Most recently, Liu et al.~\cite{liu2024adversarial} proposed the first adversarial attack framework for TS-LLMs, highlighting the need for dedicated defenses.

Defensive strategies can be grouped into model modification, input transformation and adversarial detection. Model-level defenses primarily involve adversarial training~\cite{madry2018towards,wang2024revisiting,sabir2023interpretability,bespalov2024towards,zhang2024text} or architectural adjustments such as outlier classes~\cite{grosse2017statistical}. These approaches increase computational cost, require access to attack examples and often degrade performance on benign inputs, which makes them impractical for LLMs that are expensive to retrain. Input-transformation defenses apply operations such as compression~\cite{liu2019feature} or bit-depth reduction~\cite{guo2017countering,xu2017feature}, or reconstruct data through generative models~\cite{zhou2023eliminating}. While these techniques can suppress perturbations, they frequently reduce utility and are modality specific. Detection-based defenses analyze neighborhood characteristics~\cite{ma2018characterizing} or train separate detectors~\cite{ho2022disco}, while others monitor internal layer consistency~\cite{zhang2023lsd,ma2019nic} or examine prediction stability under input modifications~\cite{xu2017feature,wang2023addition}. However, many detectors rely on adversarial examples during training and are susceptible to unseen attacks. Existing LLM-oriented defenses mostly target harmful-content filtering~\cite{kumar2023certifying,xu2022exploring,phute2023llm} and do not address vulnerabilities in numerical time-series forecasting.

\section{Overview and Design of \name}

In this section, we first present a practical threat model for adversarial attacks in energy domain, then introduce our solution \name to flag the adversarial samples.

\subsection{Threat Model}\label{sec_threat_model}

The attacker may operate in either a black-box or white-box setting. As noted in recent work on TS-LLMs~\cite{liu2024adversarial}, the black-box setting is more realistic because commercial TS-LLMs such as TimeGPT are accessible only through APIs. White-box access becomes feasible when TS-LLMs are deployed locally, for example through open-source models on Hugging Face or internal fine-tuned models maintained by utilities or aggregators, where model weights, embeddings and tokenization schemes are fully exposed. Since existing AE attacks on TS-LLMs have been studied exclusively under black-box assumptions, we primarily focus on black-box attacks and include white-box evaluations (Section~\ref{sec_whichboxAE}) for completeness.

This threat model is consistent with practical energy forecasting pipelines, where TS-LLM inputs often originate from external or partially trusted sources such as smart meters, IoT sensors, SCADA streams and third-party data feeds. Field evidence also shows that inverter telemetry and settings interfaces have exhibited software and configuration vulnerabilities in real deployments, allowing their measurement outputs to be altered or spoofed~\cite{yang2024rethink,SidEnergy,barua2020hall}. These channels are known to be susceptible to noise, tampering and data-integrity issues, which makes input-level perturbations a realistic attack surface even without model access. 

The defender refers to the entity operating the TS-LLM forecasting system, such as a grid operator, an energy aggregator, a forecasting service provider, or even a local end user. The defender is assumed to have no prior knowledge of the adversarial example (AE) strategy and to possess only a small portion of benign data for setting and updating detection thresholds, consistent with prior AE detection work. The goal is to identify adversarial inputs while minimizing false positives. \name is implemented as a lightweight plug-in module that neither modifies nor retrains the protected TS-LLM, ensuring compatibility with both commercial and open-source forecasting workflows.

\subsection{Overview}\label{sec_overview}

\begin{figure*}[t]
    \centering
    \includegraphics[width=\linewidth]{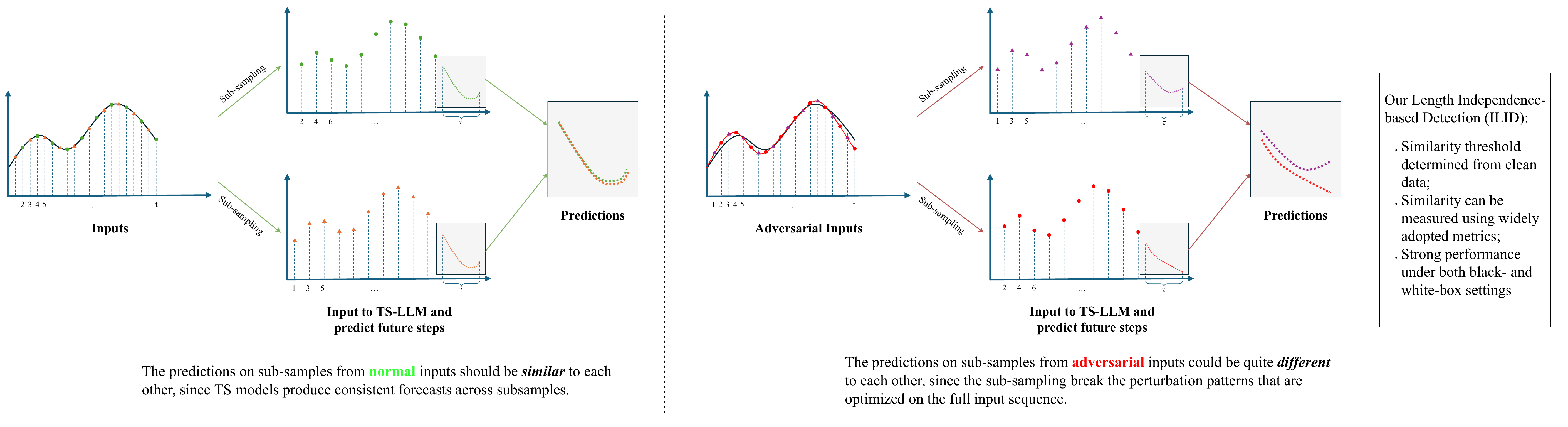}
    \caption{Overview.}
    \label{fig:Overview}
\end{figure*}

The design of \name is illustrated in Figure~\ref{fig:Overview}. It is motivated by two key insights: the nature of adversarial example (AE) attacks and the unique characteristics of TS-LLMs. The first insight is that adversarial perturbations are highly input-dependent. Since they are carefully optimized with respect to specific inputs, the perturbations are tailored to reliably mislead the target model. The second insight concerns the inherent flexibility of TS-LLMs in handling input lengths. Unlike traditional forecasting models that require fixed input sizes, TS-LLMs can naturally accommodate variable input lengths. This property forms the foundation of \name, enabling efficiency, effectiveness, and plug-in advantages in its design.

\name operates in two phases: an offline phase and an online phase. In the offline phase, a small clean dataset $\mathcal{X}$ is solely used to determine a similarity threshold, which is later employed for detecting AE inputs. Unlike a static threshold, \name supports dynamic updates as detection proceeds. Since the online phase is straightforward, we focus below on the offline phase, with implementation details deferred to Section~\ref{sec_implementation}.

In the offline phase, the first step is to subsample each clean input $\textbf{x}$ of length $T$, producing multiple subsamples $\textbf{x}_{\rm subi}$, $i \in (1,S)$, each with a reduced size (e.g., $T/2$). Next, these subsamples are fed into the underlying TS-LLM for inference, yielding predictions $\textbf{y}_{\rm subi}$. Note that the length of $\textbf{y}_{\rm subi}$ can be the same or different from the length of $\textbf{y}$ (the prediction corresponding to $\textbf{x}$), but the outputs $\textbf{y}_{\rm subi}$ are of the same length across $i \in (1,S)$. The third step is to compute the pairwise similarity among $\textbf{y}_{\rm subi}$. For benign inputs, these predictions are expected to remain consistent with each other and with $\textbf{y}$, since subsampling should not alter the temporal trend or forecasting tendency.

\name applies these steps across all clean samples, producing a distribution of similarity scores. Based on this distribution and a predefined false rejection rate (FRR), \name determines a threshold (Section~\ref{sec_implementation}) for online detection. During deployment, if an incoming input produces a similarity below this threshold, it is flagged as adversarial; otherwise, it is treated as benign.

\subsection{Implementation}\label{sec_implementation}

To be somewhat formal, following~\cite{liu2024adversarial}, we denote a time series at time $t$ as $\textbf{x}_t$, where $\textbf{x}_{t,i}$ is the i-th component of this time series. Given a recent $T$ historical observations $\textbf{x}_{t-T+1:t}$, the TS-LLM predicts the future values for the subsequet $\tau$ time steps. The prediction is expressed as: 
\begin{equation}
    \Tilde{\textbf{y}}_{t+1:t+\tau} = \textsf{TS-LLM}(\textbf{x}_{t-T+1:t}).
\end{equation}
Note that the prediction horizon is typically less than or equal to the historical horizon, that is $\tau \leq T$. In following descriptions, we exchangeable use $\textbf{x}$ with $\textbf{x}_t$ and use $\textbf{y}$ with $\Tilde{\textbf{y}}$ for simplicity.

\subsubsection{Subsampling} To sample the \textbf{x}, the sampling strategy can be diverse or flexible. One simple sampling setting is to take one component in \textbf{x} for every one component, similarly, to take two components for every two components. To ease understanding, we take the first setting. Upon this setting, we have two subsamples 
\[
\textbf{x}_{\rm sub, 1} = \{\textbf{x}_{t-T+1}, \textbf{x}_{t-T+3}, \textbf{x}_{t-T+5}, \dots, \textbf{x}_{t-3}, \textbf{x}_{t-1}\}
\]
and
\[
\textbf{x}_{\rm sub, 2} = \{\textbf{x}_{t-T+2}, \textbf{x}_{t-T+4}, \textbf{x}_{t-T+6}, \dots, \textbf{x}_{t-2}, \textbf{x}_{t}\}.
\]
Supposing that $T$ is even, $\textbf{x}_{\rm sub, 1}$ corresponds to all odd components in \textbf{x}, while $\textbf{x}_{\rm sub, 2}$ corresponds to all even components in \textbf{x}.

We emphasize that the above example is only one sampling setting, while the sampling settings can vary according to the need. In addition, the components in each subsample do not necessarily need to be non-overlapping, as in the above example. More generally, the user can randomly take a fraction (e.g., 60\%) of components from \textbf{x} to form a subsample. Such inherent randomness in \name enables it to defeat potential adaptive attacks, as we evaluate in Section~\ref{sec_sampling}.

\subsubsection{Similarity} Given each subsample $\textbf{x}_{\rm sub,i}$, a prediction with $\tau$ steps $\textbf{y}_{\rm sub, i}$ is given through \textsf{TS-LLM} function. Suppose that we have $S$ subsamples per $\textbf{x}$, there will be $S$ predictions. The pairwise similarity among these $S$ predictions $\textbf{y}_{\rm sub, i}, i\in (1, S)$ is computed. If we have $N$ clean samples in $\mathcal{X}$, then the similarity of each $\textbf{x}\in \mathcal{X}$ is computed. The similarity can be measured using widely adopted metrics, including Pearson similarity, as well as similarities derived from Euclidean distance and Manhattan distance. \name is insensitive to the choice of similarity metric, as we show in Section~\ref{sec:similarity}, although we adopt Pearson similarity for extensive evaluations.

\subsubsection{Threshold} To counter unseen AE attacks, \name assumes no prior knowledge of attack strategies. The detection threshold is determined solely from the clean dataset. After computing the similarity of each clean sample in $\mathcal{X}$, we obtain $N$ similarity values. We reasonably assume these values follow a normal distribution, from which the mean and standard deviation of the similarity distribution of benign inputs can be estimated. A false rejection rate (FRR), e.g., 1\%, that can be tolerated is then preset. The FRR is the probability that a benign input is falsely regarded as adversarial, thus rejected. The corresponding percentile of the normal distribution is calculated and chosen as the detection threshold. 

As an alternative, one may sort the $N$ similarity values in ascending order and select the $(i+1)$-th smallest value in the sorted list as the threshold, such that $\frac{i}{N} = \text{FRR}$ (e.g., FRR = 1\%). During online detection, if the similarity of a given input is lower than this threshold, the input is regarded as adversarial; otherwise, it is treated as benign.

\section{Experimental Evaluation and Results} 

In this section, we first describe the experimental setup, including the datasets and TS models used in our evaluation, then provide experimental results to showcase our detection performance against adversarial attacks.

\subsection{Setup}\label{sec_setup}

In the context of energy applications, we identify three forecasting datasets for comprehensive evaluation, each reflecting practical scenarios in electricity load, generation, and consumption forecasting. Three TS-LLMs, namely TimeGPT, Time-LLM, and TimesFM, are considered. 
TimeGPT is a proprietary Nixtla model that requires payment per query, with limited free trials upon registration, and is available only through a black-box API. Although TimeLLM is open-source, its NeuralForecast library integration allows it to be used in a similar black-box fashion. 

\subsubsection{Datasets} 
We employ three energy application-specific datasets: Electricity Transformer Temperature Hourly (ETTh2), Hourly Energy Consumption Data (NI), and Hourly Electricity Consumption and Production (Consumption).

\noindent$\bullet$ \textbf{ETTh2}~\cite{zhou2021informer} contains different subsets for long-sequence time-series forecasting (LSTF). ETTh2 is one such subset, recorded at 1-hour resolution. Each entry includes the target variable ``oil temperature'' and six power load features. The dataset spans two years and is collected from two counties in China.

\noindent$\bullet$ \textbf{NI}~\cite{PJM} provides hourly energy consumption measurements in megawatts (MW), obtained from the official PJM repository. The Northern Illinois Hub (NI) is part of PJM Interconnection, a major regional transmission organization (RTO) in the United States. In our study, the NI dataset is used to examine regional power consumption patterns.

\noindent$\bullet$ \textbf{Consumption}~\cite{Romania} contains over six years of hourly electricity consumption and production data in Romania. It reports detailed generation by source—including nuclear, wind, solar, and others—offering insights into the country’s diverse energy mix. All values are measured in megawatts (MW).

\subsubsection{TS-LLMs} 
Three TS-LLMs are employed for evaluation. TimeGPT and TimesFM perform zero-shot only forecasting on the above datasets.

\noindent$\bullet$ \textbf{TimeGPT}~\cite{garza2023timegpt} is a commercial TS-LLM trained from scratch on a vast corpus of time series data. It is explicitly designed for time series forecasting, achieving high accuracy across diverse real-world applications. TimeGPT requires a subscription for usage. For example, the \$99/month standard plan permits 10,000 API calls per month, which we have subscribed to for experiments.

\noindent$\bullet$ \textbf{Time-LLM}~\cite{jin2023time} reprograms a pretrained LLM for general time series forecasting while keeping the language model, originally trained on textual corpora, unchanged. A key component is the Prompt-as-Prefix (PaP) strategy, which improves contextual understanding and guides the model toward accurate forecasting on reformatted time series data. To be user-friendly, Time-LLM has incorporated this PaP into its system prompt, where the user does not need to provide and can use the Time-LLM for forecasting. Time-LLM is integrated into the \texttt{NeuralForecast} library, making it straightforward to use in practice.

\noindent$\bullet$ \textbf{TimesFM}~\cite{das2024decoder} is a pretrained time-series forecasting foundation model developed by Google Research, trained on 100 billion real-world time points. The public model from Hugging Face is able to be loaded and used directly, without additional fine-tuning. TimesFM-2.0-500M is used in our experiment.

\subsection{Adversarial Example Attack} 
There is only one recent adversarial example (AE) attack, namely Directional Gradient Approximation (DGA)~\cite{liu2024adversarial}, specifically devised against TS-LLMs in a realistic black-box setting. Therefore, our evaluations primarily focus on this attack, while we defer evaluations of white-box attacks such as FGSM, BIM, and PGD to Section~\ref{sec_whichboxAE}. The forecasting performance of TS-LLMs is assessed using three metrics: Mean Absolute Error (MAE), Mean Squared Error (MSE), and the $R^2$ score. The objective of an AE attack is to deviate the prediction from its normal output, such that MAE and MSE increase notably, while $R^2$ decreases. Note that if the performance deviation is marginal, the AE attack cannot be regarded as successful.

We apply DGA to all three datasets against each of the three TS-LLMs. Table~\ref{tab:foundation_models_full} summarizes TS-LLM performance under both clean and adversarial example queries. We ensure that the attacks succeed before evaluating the defensive effect of \name. As expected, both MAE and MSE exhibit a marked increase under AE queries compared to clean sample queries, whereas the $R^2$ score demonstrates a noticeable decline as expected.

\begin{table*}[t]
\centering
\caption{Forecasting performance of foundation models under clean and adversarial settings.}
\label{tab:foundation_models_full}
\resizebox{\linewidth}{!}{
\begin{tabular}{llcccccccccccccccccc}
\toprule
\multirow{2}{*}{Dataset} & \multirow{2}{*}{Target} 
& \multicolumn{3}{c}{TimeGPT (Clean)} 
& \multicolumn{3}{c}{TimeGPT (AE)}
& \multicolumn{3}{c}{TimeLLM (Clean)} 
& \multicolumn{3}{c}{TimeLLM (AE)} 
& \multicolumn{3}{c}{TimesFM (Clean)} 
& \multicolumn{3}{c}{TimesFM (AE)} 
\\
\cmidrule(rl){3-5}\cmidrule(rl){6-8}\cmidrule(rl){9-11}\cmidrule(rl){12-14}\cmidrule(rl){15-17}\cmidrule(rl){18-20}
& 
& MAE & MSE & $\text{R}^2$
& MAE & MSE & $\text{R}^2$
& MAE & MSE & $\text{R}^2$
& MAE & MSE & $\text{R}^2$
& MAE & MSE & $\text{R}^2$
& MAE & MSE & $\text{R}^2$
\\
\midrule
ETTh2 & OT
& 0.269 & 0.128 & 87.1\%
& 0.345 & 0.203 & 79.6\%
& 0.257 & 0.124 & 82.7\%
& 0.368 & 0.237 & 66.9\%
& 0.245 & 0.118 & 88.8\%
& 0.324 & 0.181 & 82.8\%
\\
NI & NI-MW
& 0.282 & 0.196 & 78.4\%
& 0.448 & 0.351 & 61.4\%
& 0.480 & 0.414 & 71.0\%
& 0.555 & 0.533 & 62.7\%
& 0.257 & 0.182 & 77.4\%
& 0.425 & 0.330 & 59.0\%
\\
Consumption & Consumption
& 0.154 & 0.050 & 93.6\%
& 0.423 & 0.284 & 63.4\%
& 0.403 & 0.293 & 69.4\%
& 0.534 & 0.477 & 50.3\%
& 0.208 & 0.107 & 88.3\%
& 0.397 & 0.269 & 70.5\%
\\
\bottomrule
\end{tabular}
}
\end{table*}

\subsection{Offline Similarity Threshold}\label{sec_offline_similarity_th} 

To be comprehensive, we employ different sampling strategies to generate subsamples with three subsample sizes: 240, 160, and 120 time steps (input length), given the same query input \textbf{x} with $T = 480$ time steps. For the 240-step sampling, we take all odd and even components in \textbf{x}, respectively, as subsamples, resulting in two subsamples per input sample under this setting. For the 160-step sampling, we subsample \textbf{x} by selecting one element every three steps (i.e., with a stride of 3), respectively, as subsamples, resulting in three subsamples per input sample under this setting. For the 120-step sampling, subsamples are generated from input \textbf{x} by selecting one element every four time steps, corresponding to a stride of 4, resulting in four subsamples per input sample under this setting.

\begin{figure*}[t]
\centering
\begin{subfigure}[t]{0.3\linewidth}
\includegraphics[width=\linewidth]{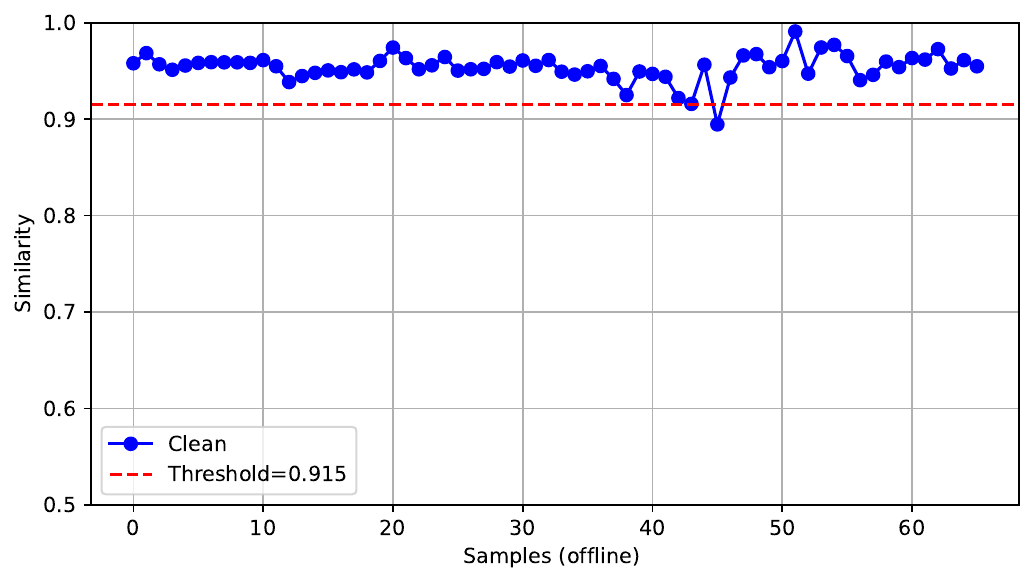}
\caption{Offline: subsample length = 240 (ETTh2)}
\label{fig_Etth2-240-offline}
\end{subfigure}
\begin{subfigure}[t]{0.3\linewidth}
\includegraphics[width=\linewidth]{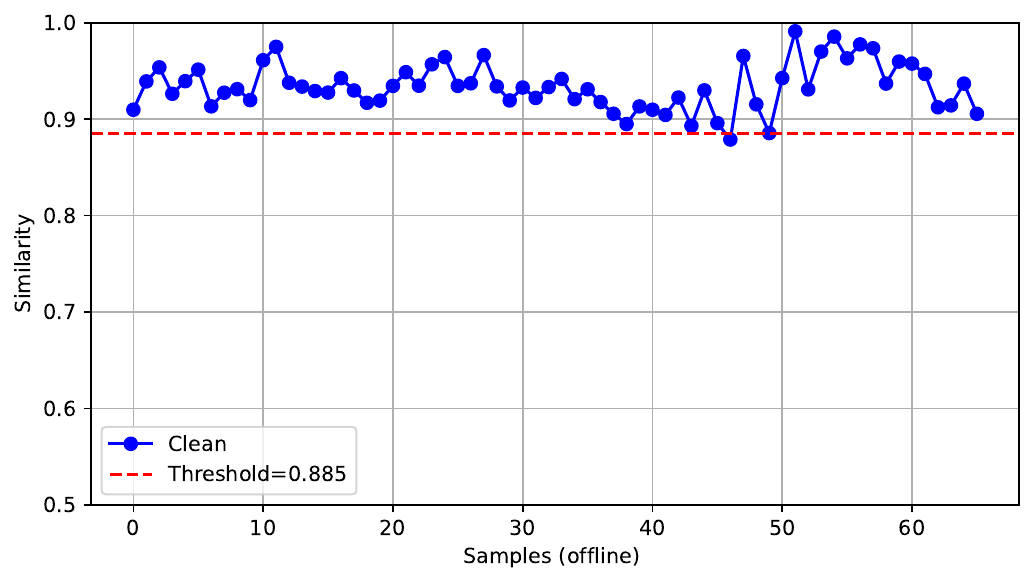}
\caption{Offline: subsample length = 160 (ETTh2)}
\label{fig_Etth2-160-offline}
\end{subfigure}
\begin{subfigure}[t]{0.3\linewidth}
\includegraphics[width=\linewidth]{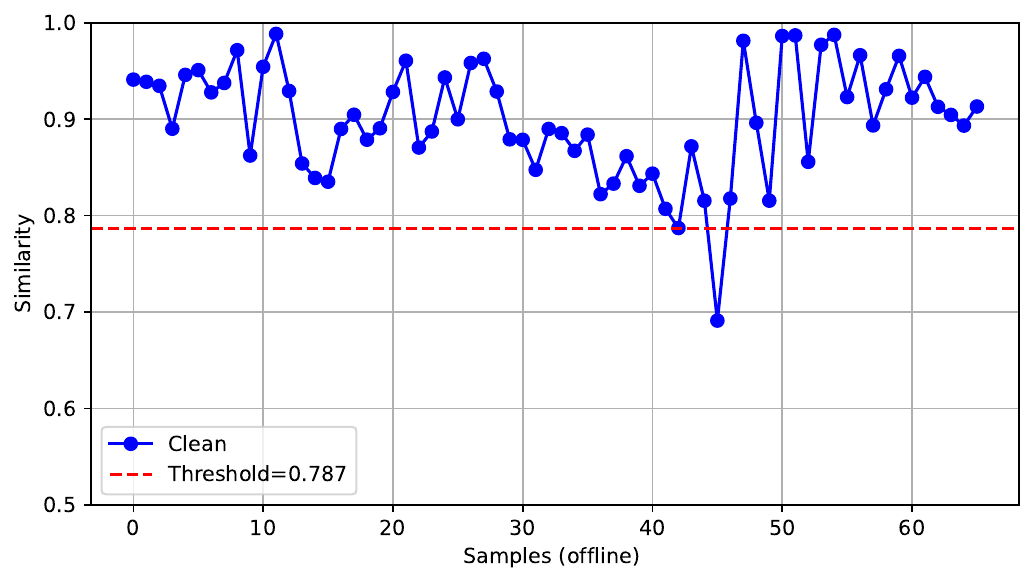}
\caption{Offline: subsample length = 120 (ETTh2)}
\label{fig_Etth2-120-offline}
\end{subfigure}

\vspace{0.3em} 

\begin{subfigure}[t]{0.3\linewidth}
\includegraphics[width=\linewidth]{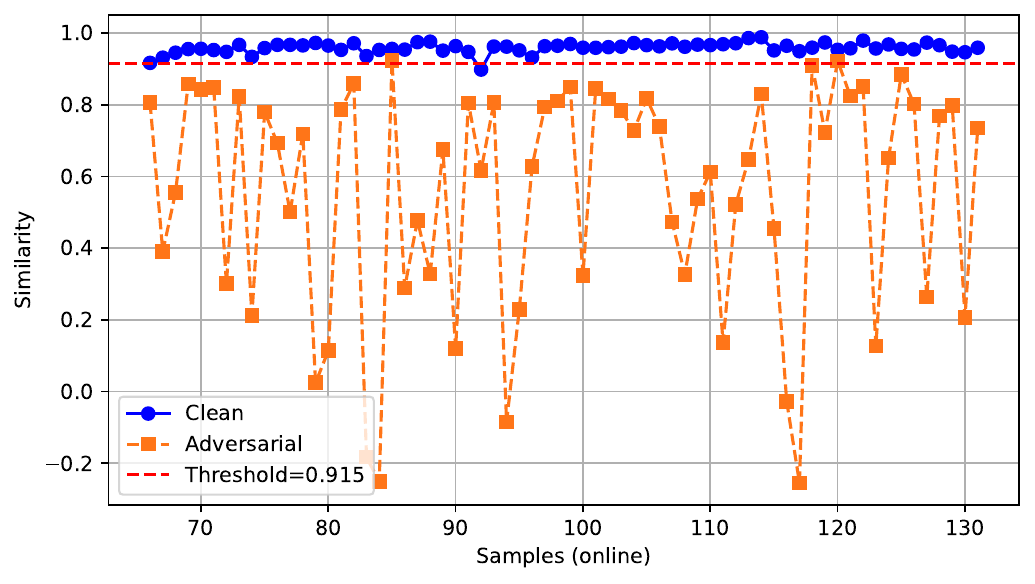}
\caption{Online: subsample length = 240 (ETTh2)}
\label{fig_Etth2-240-onlineGPT}
\end{subfigure}
\begin{subfigure}[t]{0.3\linewidth}
\includegraphics[width=\linewidth]{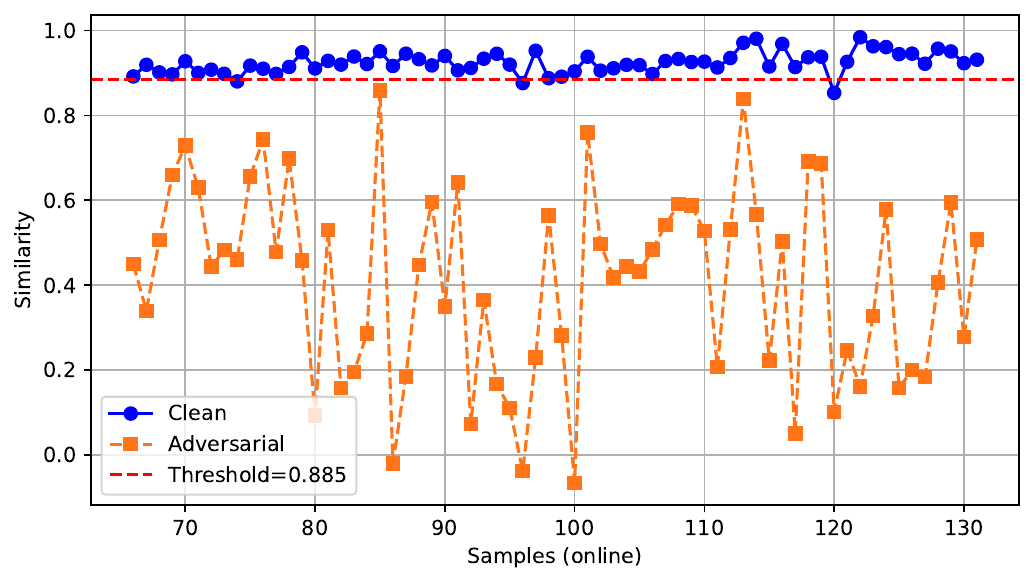}
\caption{Online: subsample length = 160 (ETTh2)}
\label{fig_Etth2-160-onlineGPT}
\end{subfigure}
\begin{subfigure}[t]{0.3\linewidth}
\includegraphics[width=\linewidth]{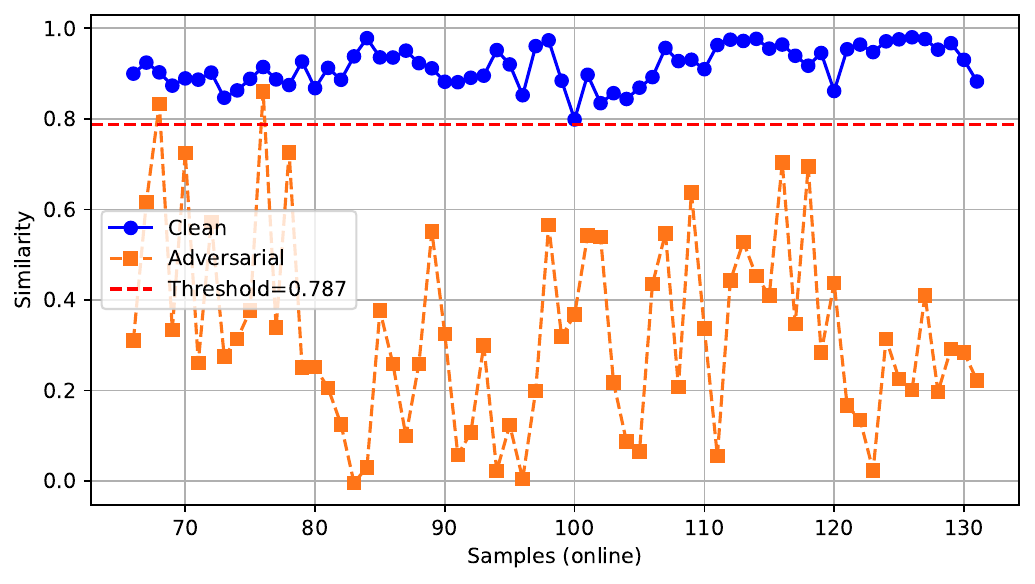}
\caption{Online: subsample length = 120 (ETTh2)}
\label{fig_Etth2-120-onlineGPT}
\end{subfigure}

\caption{Offline (top) and online (bottom) detection results for ETTh2 using TimeGPT with different subsample lengths.}
\label{fig:Etth2GPT-offline-online}
\end{figure*}

Afterward, the pairwise similarity of the forecasting values from subsamples of the same size is computed and then used to determine the threshold based on the similarities of all retained clean samples in $\mathcal{X}$. The similarity distributions for clean queries under subsample lengths of 240, 160 and 120 in the ETTh2 dataset are shown in Figure~\ref{fig:Etth2GPT-offline-online} (top row). Here, the smallest mean pairwise similarity value is tolerated, while the second-smallest value is regarded as the threshold. More specifically, in Figure~\ref{fig_Etth2-240-offline}, the threshold is set to 0.915, corresponding to tolerating one false positive. 

It is worth noting that this threshold, although initially determined offline, does not need to remain static; it can be updated dynamically during online detection, as detailed in Section~\ref{sec:dynamic}.

\subsection{Online Detection Results}\label{sec_online_det_results} 

The dataset in all experiments is split into two equal parts; the first part is used to determine the offline threshold as detailed above, while the remaining samples are for online evaluation.

\subsubsection{TimeGPT} We evaluate TimeGPT with all three datasets: ETTh2, NI and Consumption.

\noindent$\bullet$ {\it ETTh2.} 
The total number of ETTh2 samples is 132, each with 480 timesteps. Each sample is used to forecast the next 48 timesteps. The first 66 samples are used for threshold determination, while the remaining 66 samples are employed for online evaluation.

When the subsample length is 240 timesteps, the threshold is 0.915, as shown in Figure~\ref{fig_Etth2-240-offline}. Similarly, the threshold is 0.885 in Figure~\ref{fig_Etth2-160-offline} with a subsample length of 160 timesteps, and 0.787 in Figure~\ref{fig_Etth2-120-offline} with a subsample length of 120 timesteps. In all cases, the threshold is determined by tolerating one false positive (as detailed in Section~\ref{sec_offline_similarity_th}).

The bottom subfigures in Figure~\ref{fig:Etth2GPT-offline-online} present the corresponding online detection results based on the preset thresholds. In Figure~\ref{fig_Etth2-240-onlineGPT}, with a subsample length of 240 timesteps, only one clean sample has a Pearson similarity below the threshold of 0.915 (blue), while two adversarial samples have Pearson similarities above it (orange). This indicates that one clean sample and two adversarial samples fail to be correctly detected when using the offline-determined threshold of 0.915. Figures~\ref{fig_Etth2-160-onlineGPT} and~\ref{fig_Etth2-120-onlineGPT} show the detection results when subsample lengths are 160 and 120 timesteps, respectively.

More specifically, the FRR and FAR are detailed in Table~\ref{tab:TimeGPT_detection_vertical}. The preset FRR is 1.5\% across all subsample lengths. For online detection, when the subsample length is 240 timesteps, the online FRR for clean samples remains at 1.5\%, while the FAR for adversarial samples is 3.0\%. With a subsample length of 160 timesteps, the online FRR increases to 4.5\%, while the FAR is 0\%. For a subsample length of 120 timesteps, the FRR is 0\% and the FAR rises to 3.0\%. The slight discrepancy between the preset FRR and the online FRR is potentially due to the small number of testing samples. Nevertheless, the differences are minor and acceptable.

\begin{figure*}[t]
\centering
\begin{subfigure}[t]{0.3\linewidth}
\includegraphics[width=\linewidth]{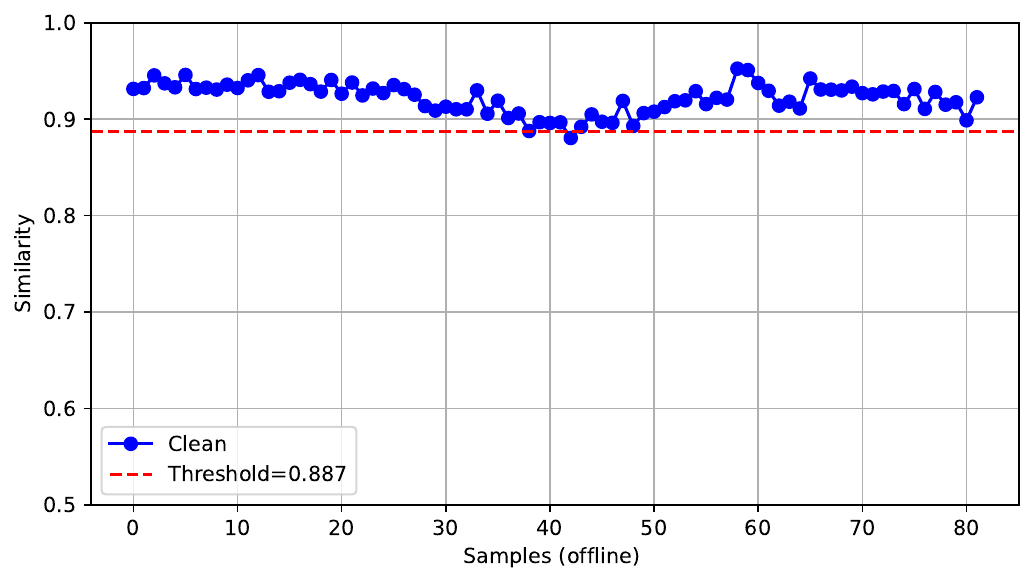}
\caption{Offline: input length = 360 (Consum.)}
\label{fig_Consumption-360-offline}
\end{subfigure}
\begin{subfigure}[t]{0.3\linewidth}
\includegraphics[width=\linewidth]{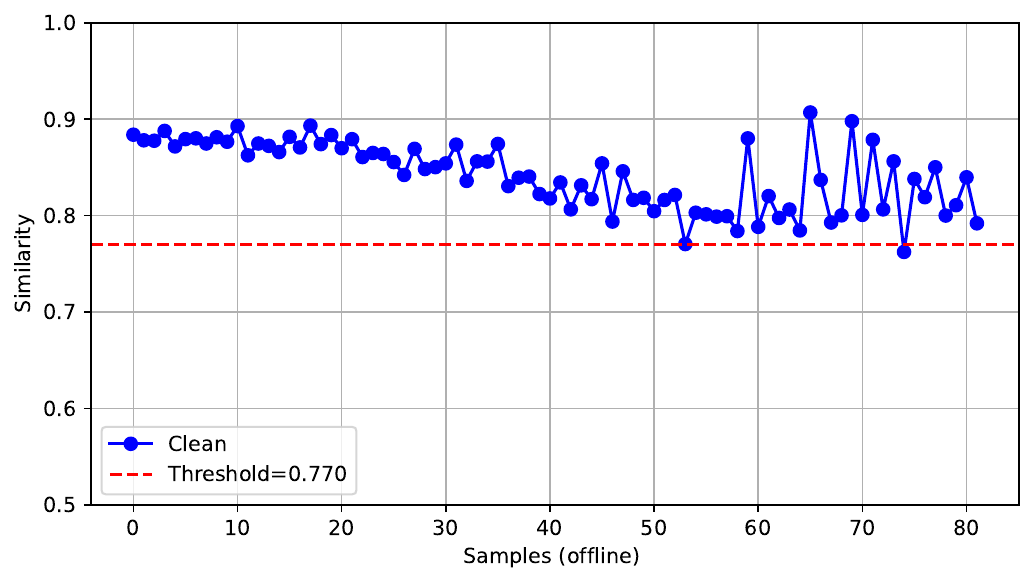}
\caption{Offline: input length = 240 (Consum.)}
\label{fig_Consumption-240-offline}
\end{subfigure}
\begin{subfigure}[t]{0.3\linewidth}
\includegraphics[width=\linewidth]{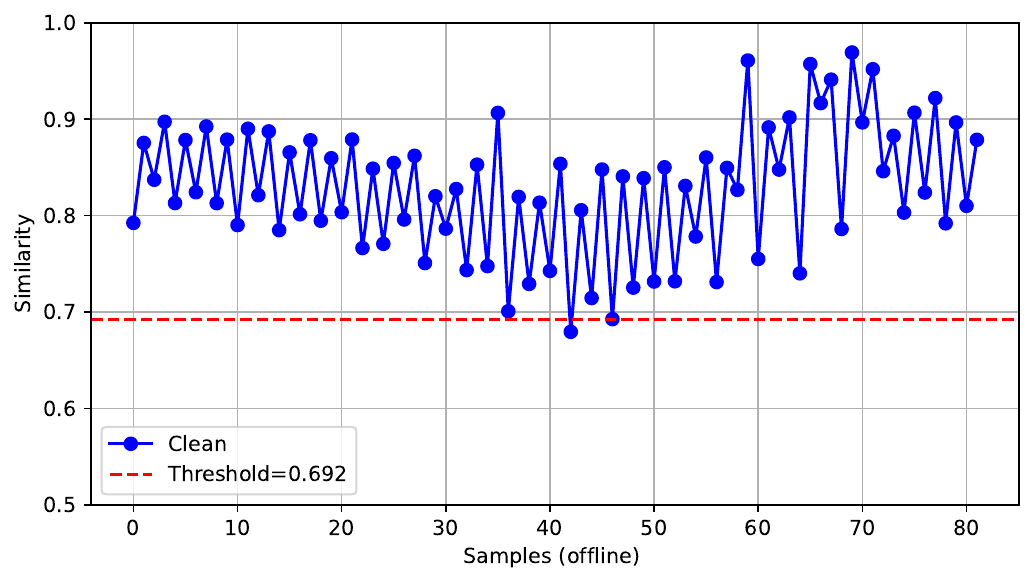}
\caption{Offline: input length = 180 (Consum.)}
\label{fig_Consumption-180-offline}
\end{subfigure}

\vspace{0.3em}

\begin{subfigure}[t]{0.3\linewidth}
\includegraphics[width=\linewidth]{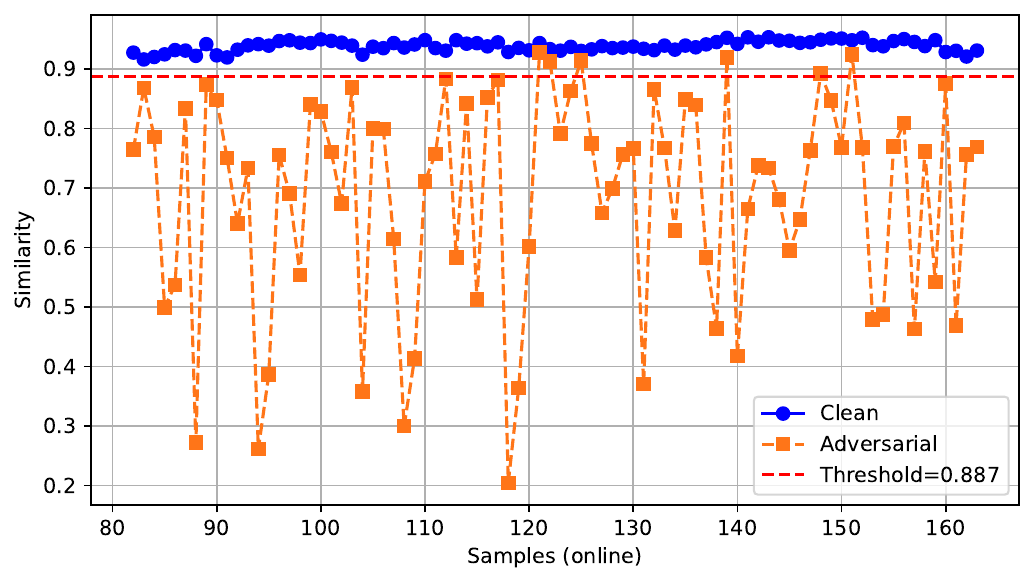}
\caption{Online: input length = 360 (Consum.)}
\label{fig_Consumption-360-online}
\end{subfigure}
\begin{subfigure}[t]{0.3\linewidth}
\includegraphics[width=\linewidth]{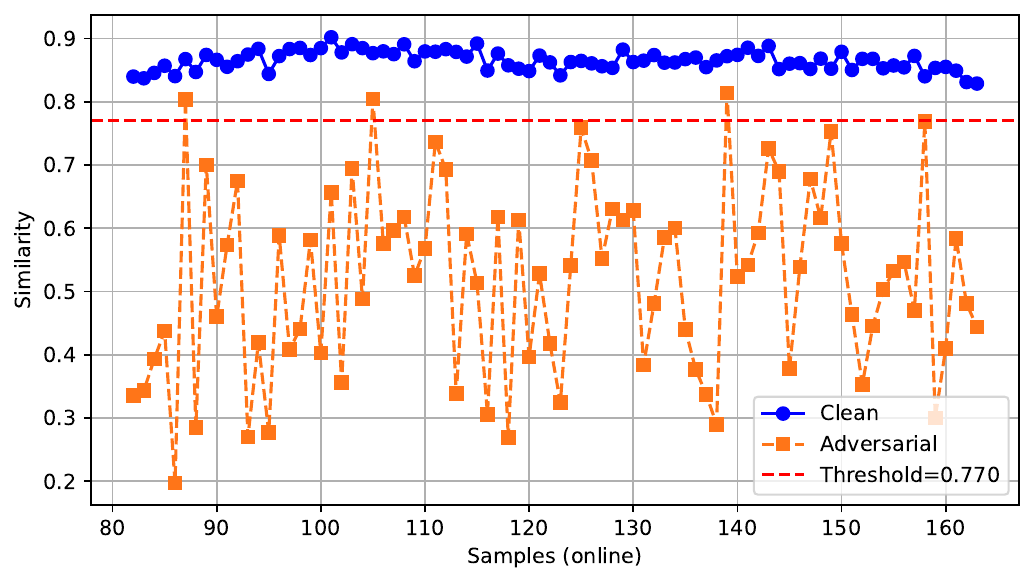}
\caption{Online: input length = 240 (Consum.)}
\label{fig_Consumption-240-online}
\end{subfigure}
\begin{subfigure}[t]{0.3\linewidth}
\includegraphics[width=\linewidth]{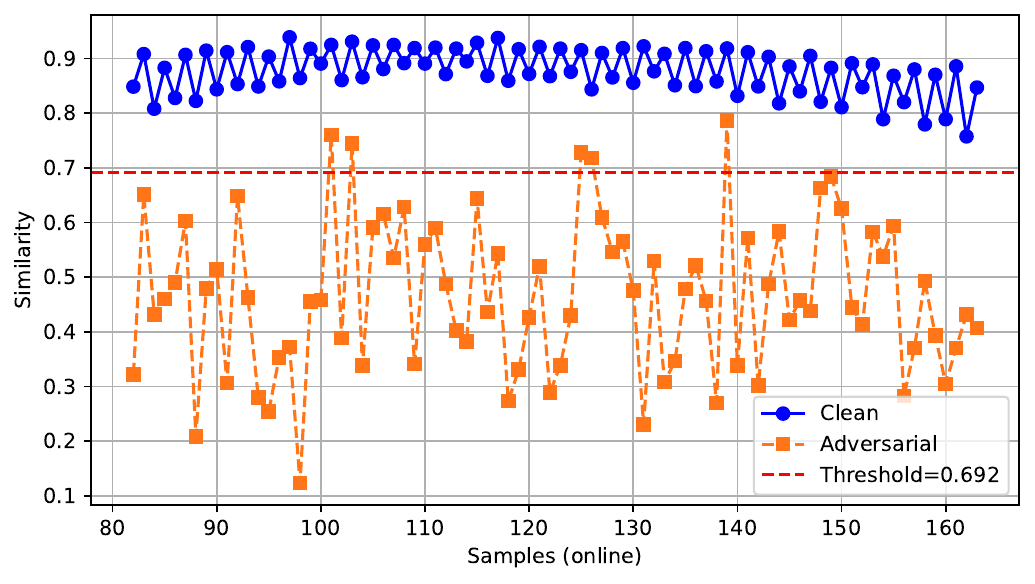}
\caption{Online: input length = 180 (Consum.)}
\label{fig_Consumption-180-online}
\end{subfigure}

\caption{Offline (top) and online (bottom) detection results for Consumption using TimeGPT with different subsample lengths.}
\label{fig:Consumption-offline-online}
\end{figure*}

\noindent$\bullet$ \textit{Consumption.} The Consumption dataset consists of 164 samples, each with 720 timesteps, used to forecast the next 36 hours. As shown in Figure~\ref{fig:Consumption-offline-online}, the offline preset thresholds are 0.887, 0.770, and 0.692 when the subsample length is 360, 240, and 180 timesteps, respectively. 
The FRR and FAR results are summarized in Table~\ref{tab:TimeGPT_detection_vertical}. With an offline tolerance FRR of 1.2\% for all subsample lengths, the online FRR for clean samples is 0\% in all cases, while the FAR for adversarial samples is 7.3\% (360 timesteps), 3.7\% (240 timesteps), and 6.1\% (180 timesteps).

\begin{figure*}[t]
\centering

\begin{subfigure}[t]{0.28\linewidth}
    \includegraphics[width=\linewidth]{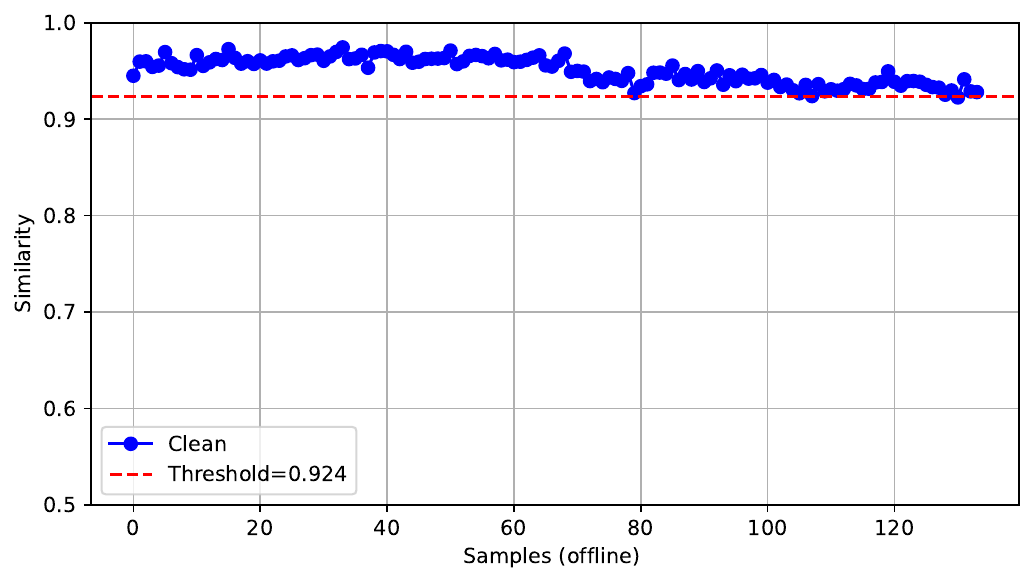}
    \caption{Offline: input length = 240 (NI)}
    \label{fig_NI-240-offline}
\end{subfigure}
\begin{subfigure}[t]{0.28\linewidth}
    \includegraphics[width=\linewidth]{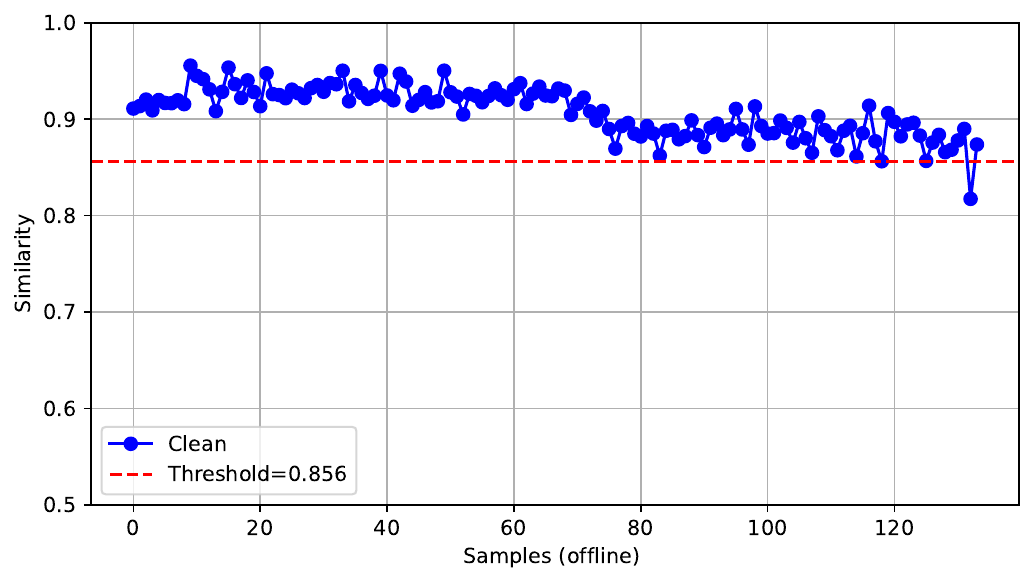}
    \caption{Offline: input length = 160 (NI)}
    \label{fig_NI-160-offline}
\end{subfigure}
\begin{subfigure}[t]{0.28\linewidth}
    \includegraphics[width=\linewidth]{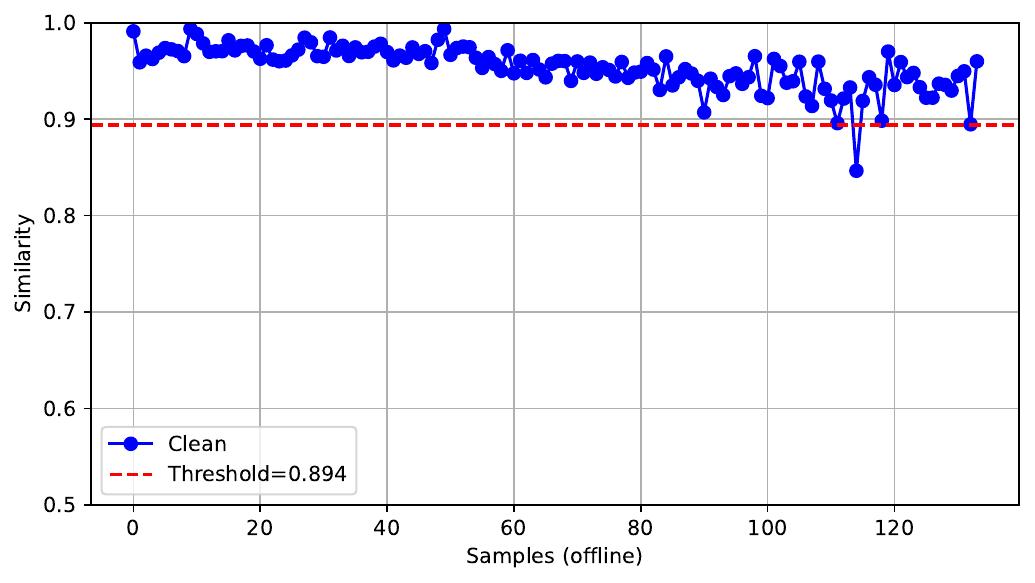}
    \caption{Offline: input length = 120 (NI)}
    \label{fig_NI-120-offline}
\end{subfigure}

\vspace{0.3em} 

\begin{subfigure}[t]{0.28\linewidth}
    \includegraphics[width=\linewidth]{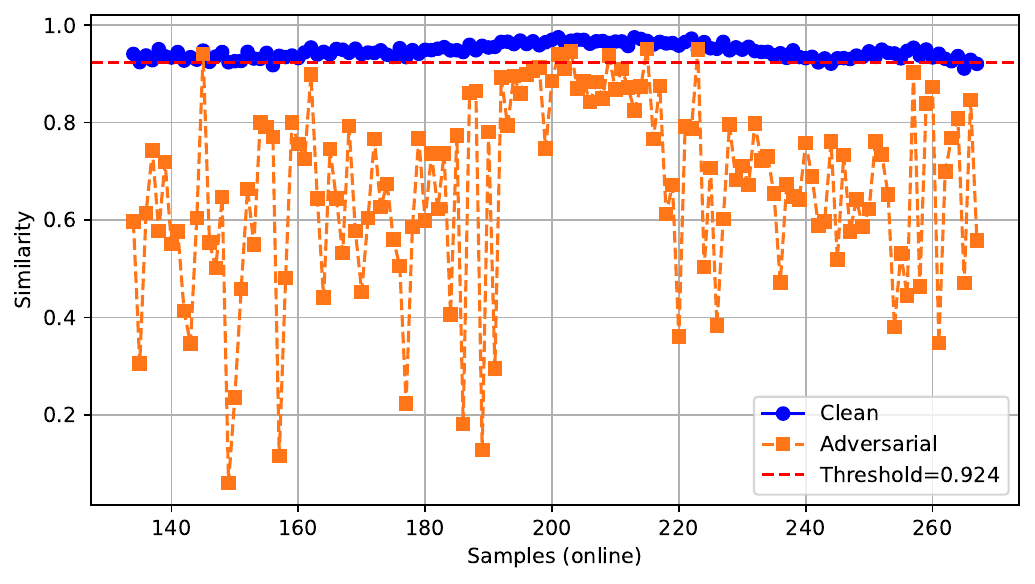}
    \caption{Online: input length = 240 (NI)}
    \label{fig_NI-240-online}
\end{subfigure}
\begin{subfigure}[t]{0.28\linewidth}
    \includegraphics[width=\linewidth]{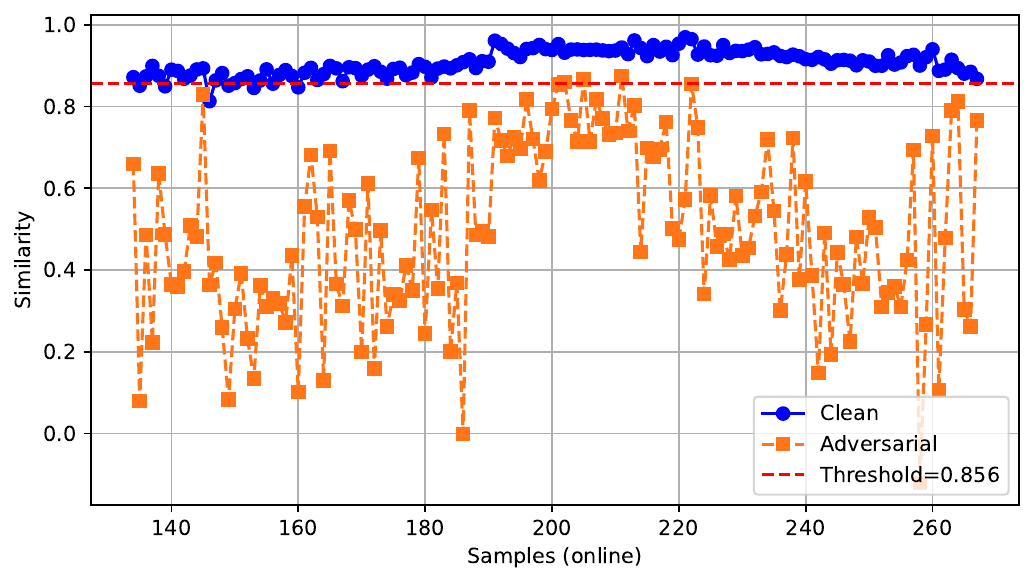}
    \caption{Online: input length = 160 (NI)}
    \label{fig_NI-160-online}
\end{subfigure}
\begin{subfigure}[t]{0.28\linewidth}
    \includegraphics[width=\linewidth]{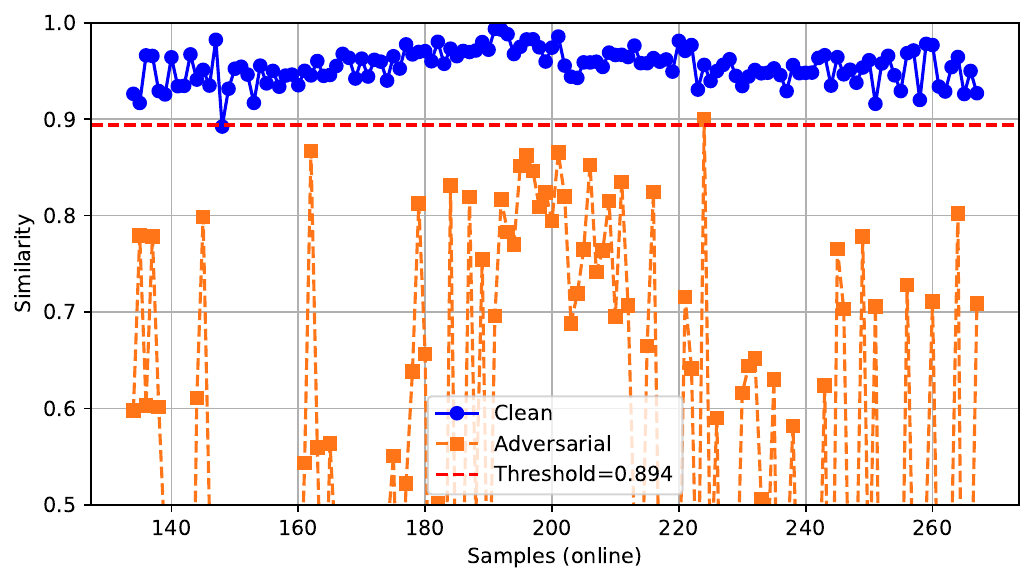}
    \caption{Online: input length = 120 (NI)}
    \label{fig_NI-120-online}
\end{subfigure}

\caption{Offline (top) and online (bottom) detection results for NI using TimeGPT with different subsample lengths.}
\label{fig:NI-offline-online}
\end{figure*}

\noindent$\bullet$ \textit{NI.} The NI dataset consists of 268 samples, each with 480 timesteps, used to forecast the subsequent 48 timesteps.  
As shown in Figure~\ref{fig:NI-offline-online}, the offline thresholds are 0.924 for a subsample length of 240, 0.856 for a subsample length of 160, and 0.894 for a subsample length of 120. 
The FRR and FAR results are summarized in Table~\ref{tab:TimeGPT_detection_vertical}. With the same preset FRR of 0.7\% across all subsample lengths, the FRR for clean samples and the FAR for adversarial samples are both 4.5\% when the subsample length is 240 timesteps. For a subsample length of 160 timesteps, the FRR for clean samples is 5.2\% and the FAR for adversarial samples is 2.2\%. For a subsample length of 120 timesteps, both the online FRR for clean samples and the FAR for adversarial samples are 0.7\%.

\begin{table}[t]
\centering
\caption{Detection performance on three datasets (ETTh2, Consumption, and NI) using TimeGPT.}
\label{tab:TimeGPT_detection_vertical}
\resizebox{.8\linewidth}{!}{
\begin{tabular}{cccccc}
\toprule
\multirow{2}{*}{Dataset}&
\multirow{2}{*}{Length}
&\multirow{2}{*}{\begin{tabular}{@{}c@{}}Threshold\end{tabular}}
& \multicolumn{1}{c}{Offline} 
& \multicolumn{2}{c}{Online}       \\
\cmidrule(r){4-4}\cmidrule(r){5-6}
& & & FRR  &FRR & FAR  \\
\midrule

\multirow{3}{*}{ETTh2}
& 240 & 0.915 & 1.5\% & 1.5\% & 3.0\% \\
& 160 & 0.885 & 1.5\% & 4.5\% & 0.0\% \\
& 120 & 0.787 & 1.5\% & 0.0\% & 3.0\% \\
\midrule

\multirow{3}{*}{Consumption}
& 360 & 0.887 & 1.2\% & 0.0\% & 7.3\% \\
& 240 & 0.770 & 1.2\% & 0.0\% & 3.7\% \\
& 180 & 0.692 & 1.2\% & 0.0\% & 6.1\% \\
\midrule

\multirow{3}{*}{NI}
& 240 & 0.924 & 0.7\% & 4.5\% & 4.5\% \\
& 160 & 0.856 & 0.7\% & 5.2\% & 2.2\% \\
& 120 & 0.894 & 0.7\% & 0.7\% & 0.7\% \\
\bottomrule

\end{tabular}}
\end{table}

\subsubsection{TimeLLM}
\begin{figure*}[t]
\centering

\begin{subfigure}[t]{0.32\linewidth}
    \includegraphics[width=\linewidth]{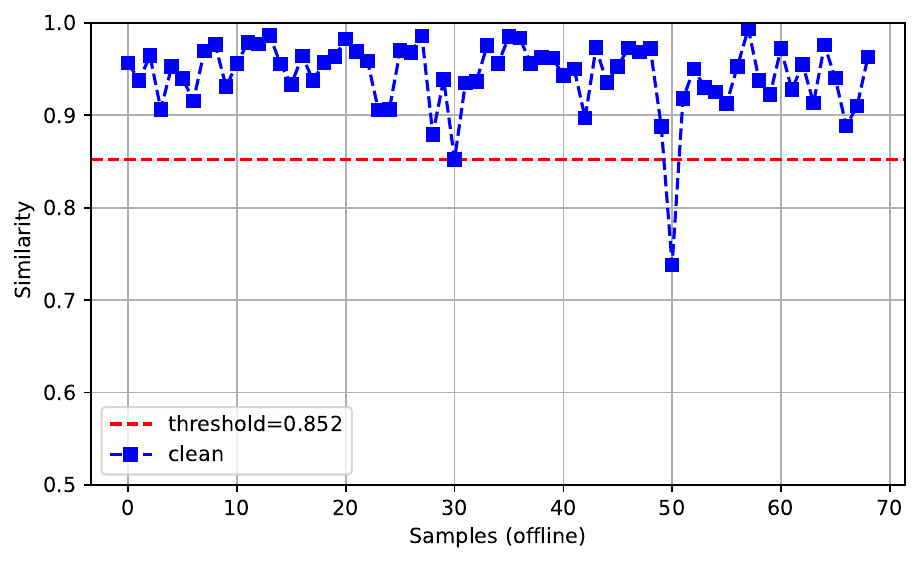}
    \caption{Offline: subsample length = 56 (ETTh2)}
    \label{fig_Etth2-56-offline}
\end{subfigure}
\begin{subfigure}[t]{0.32\linewidth}
    \includegraphics[width=\linewidth]{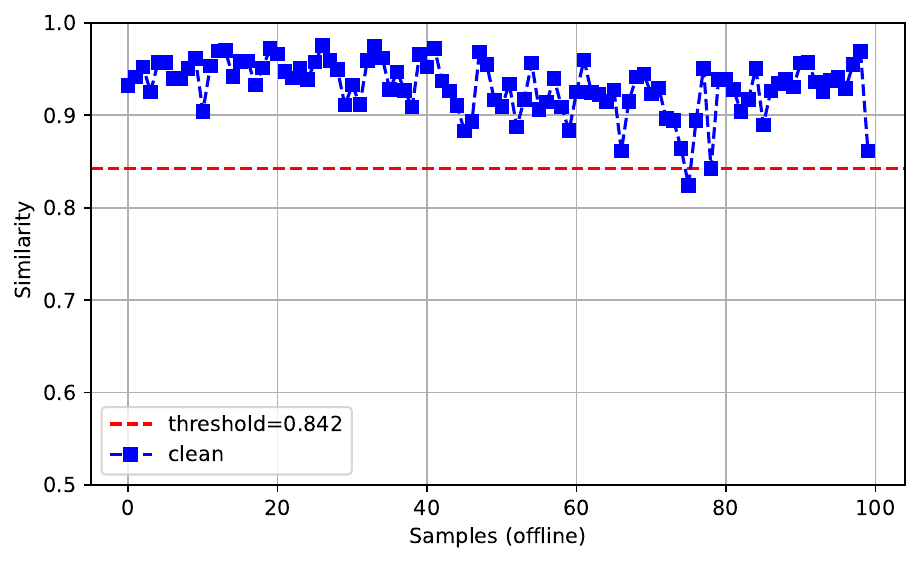}
    \caption{Offline: subsample length = 240 (Consum.)}
    \label{fig_Consumption-240-offline-TimeLLM}
\end{subfigure}
\begin{subfigure}[t]{0.32\linewidth}
    \includegraphics[width=\linewidth]{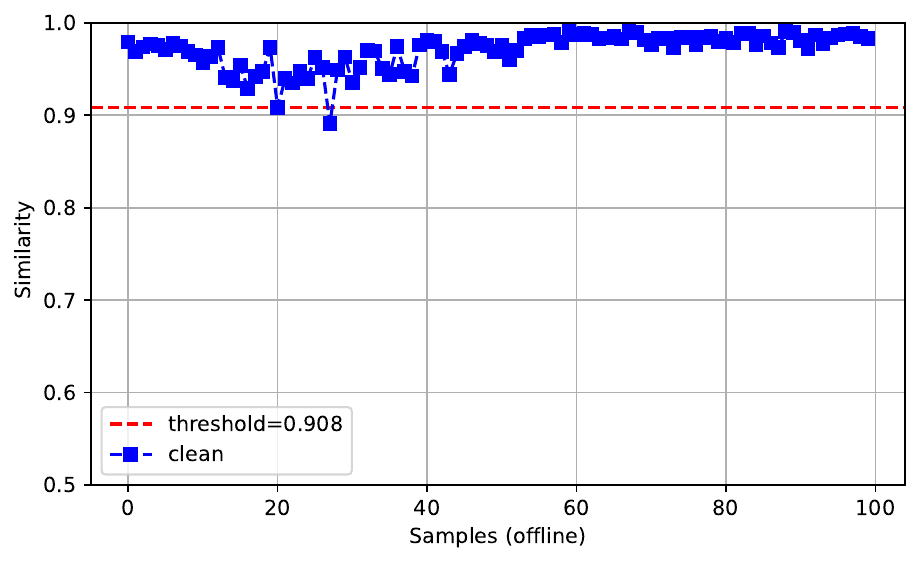}
    \caption{Offline: subsample length = 180 (NI)}
    \label{fig_NI-180-offline}
\end{subfigure}

\vspace{0.3em}

\begin{subfigure}[t]{0.32\linewidth}
    \includegraphics[width=\linewidth]{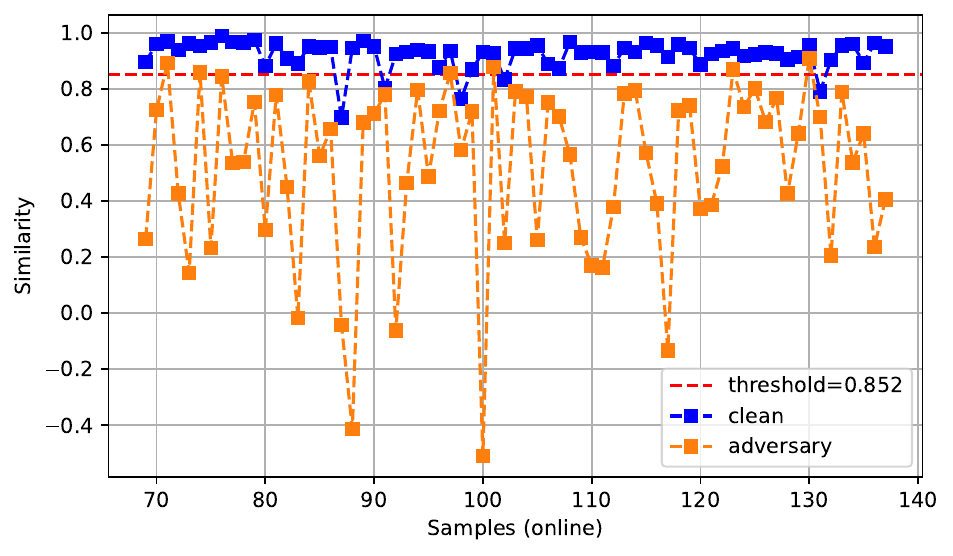}
    \caption{Online: subsample length = 56 (ETTh2)}
    \label{fig_Etth2-56-online}
\end{subfigure}
\begin{subfigure}[t]{0.32\linewidth}
    \includegraphics[width=\linewidth]{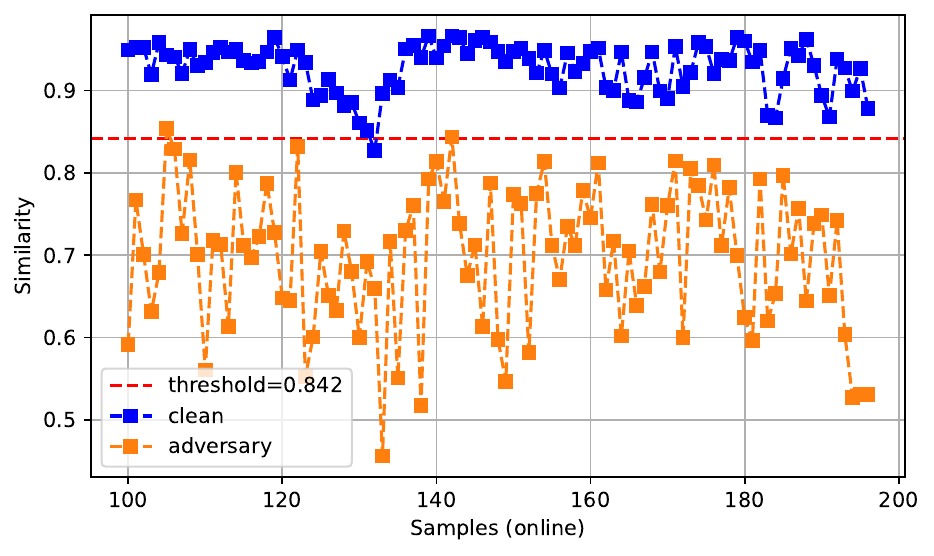}
    \caption{Online: subsample length = 240 (Consum.)}
    \label{fig_Consumption-240-online-TimeLLM}
\end{subfigure}
\begin{subfigure}[t]{0.32\linewidth}
    \includegraphics[width=\linewidth]{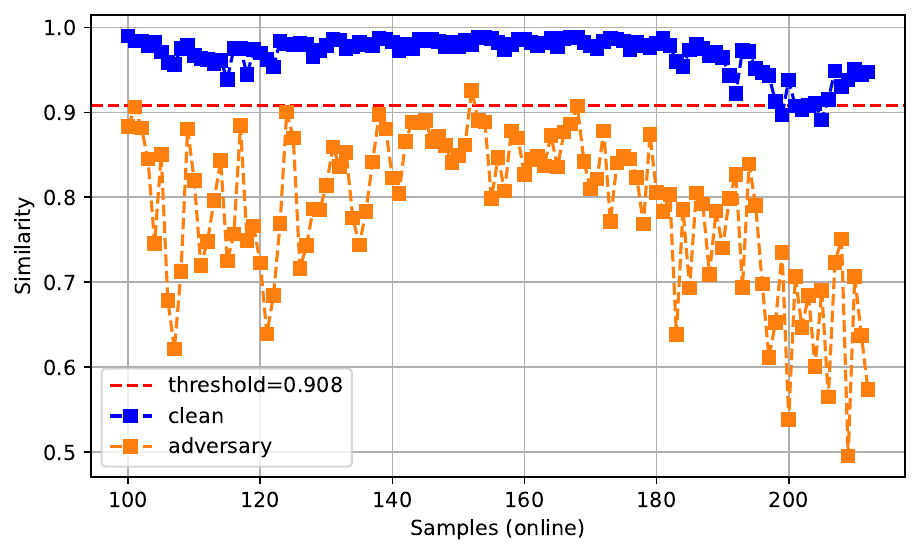}
    \caption{Online: subsample length = 180 (NI)}
    \label{fig_NI-180-online}
\end{subfigure}

\caption{Offline (top) and online (bottom) detection results for TimeLLM across different datasets.}
\label{fig:LLM-offline-online}
\end{figure*}

Figure~\ref{fig:LLM-offline-online} shows the detection performance using TimeLLM on all three datasets.  
The ETTh2 dataset consists of 138 samples, each with 168 timesteps to forecast the next 24. A subsample length of 56 is selected, with a sampling stride of three timestep. As shown in Figure~\ref{fig_Etth2-56-offline}, the offline threshold is 0.852 with a preset FRR of 1.4\%, determined using the first 69 samples. In Figure~\ref{fig_Etth2-56-online}, five clean samples fall below the threshold (blue), resulting in an online FRR of 7.2\%, while six adversarial samples fall above the threshold (orange), corresponding to a FAR of 8.7\%.  

The Consumption contains 198 samples, each with 720 timesteps to forecast the next 24. A subsample length of 240 is chosen, with a sampling stride of three timesteps. As shown in Figure~\ref{fig_Consumption-240-offline-TimeLLM}, the offline threshold is 0.842 with a preset FRR of 1.0\%. In Figure~\ref{fig_Consumption-240-online-TimeLLM}, one clean sample falls below the threshold and two adversarial samples fall above it, corresponding to an online FRR of 1.0\% and a FAR of 2.0\%.  

The NI dataset consists of 213 samples, each with 720 timesteps to forecast the following 24. A subsample length of 180 is used, with a sampling stride of four timesteps. Using the first 100 samples, the threshold is set to 0.908, as shown in Figure~\ref{fig_NI-180-offline}, with a preset FRR of 1.0\%. In Figure~\ref{fig_NI-180-online}, four clean samples and one adversarial sample fail to be correctly detected, corresponding to an online FRR of 3.5\% and a FAR of 0.9\%.  
For all three datasets, the FRR and FAR results are further summarized in Table~\ref{tab:TimeLLM}.  

\input{Table/Tab_TimeLLM}

\subsubsection{TimesFM} For this evaluation, all sample sizes of the ETTh2, Consumption, and NI datasets are set to 256 timesteps to forecast the future 48 timesteps. The subsample length is chosen to be 85 by applying stride-sampling with a stride of 2, starting respectively from the 2nd, 3rd, and 4th elements, rendering 3 subsamples.  

\begin{figure*}[t]
\centering

\begin{subfigure}[t]{0.32\linewidth}
    \includegraphics[width=\linewidth]{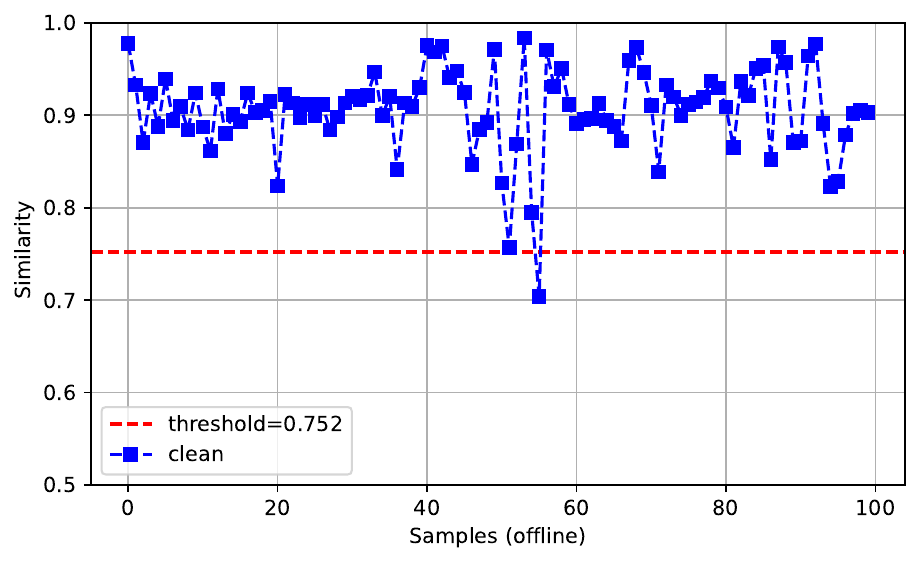}
    \caption{Offline: subsample length = 85 (ETTh2)}
    \label{fig_Etth2-85-offline234}
\end{subfigure}
\begin{subfigure}[t]{0.32\linewidth}
    \includegraphics[width=\linewidth]{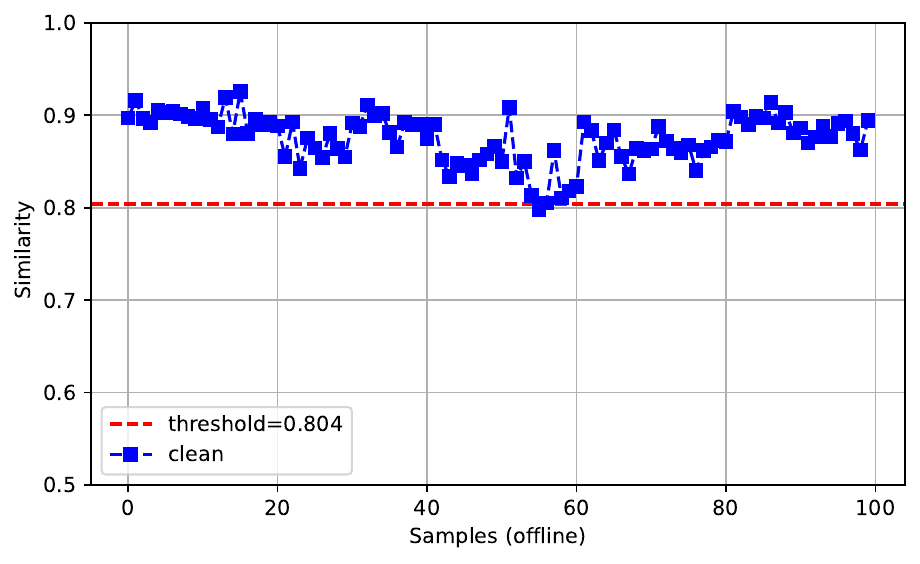}
    \caption{Offline: subsample length = 85 (Consum.)}
    \label{fig_Etth2-85-offline234.pdf}
\end{subfigure}
\begin{subfigure}[t]{0.32\linewidth}
    \includegraphics[width=\linewidth]{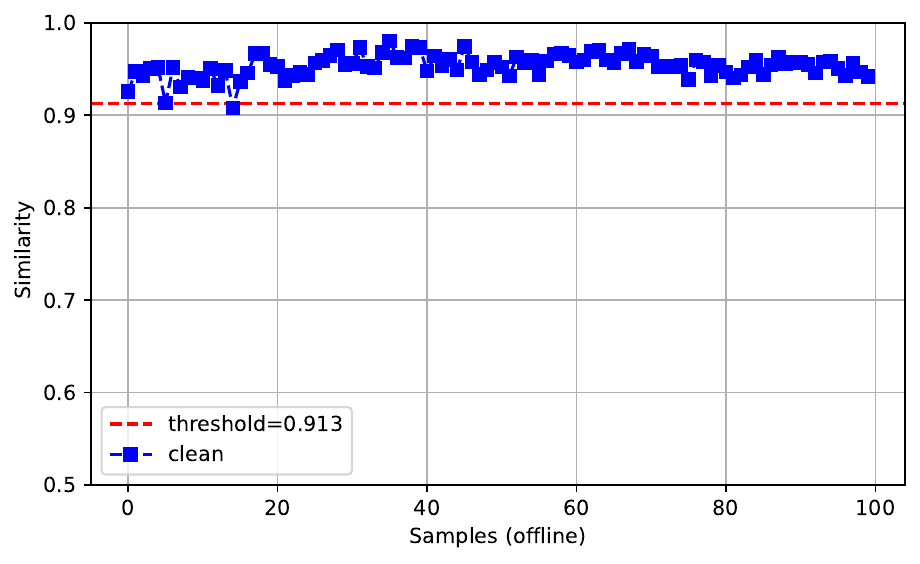}
    \caption{Offline: subsample length = 85 (NI)}
    \label{fig_NI-85-offline2-3-3-4}
\end{subfigure}

\vspace{0.3em} %

\begin{subfigure}[t]{0.32\linewidth}
    \includegraphics[width=\linewidth]{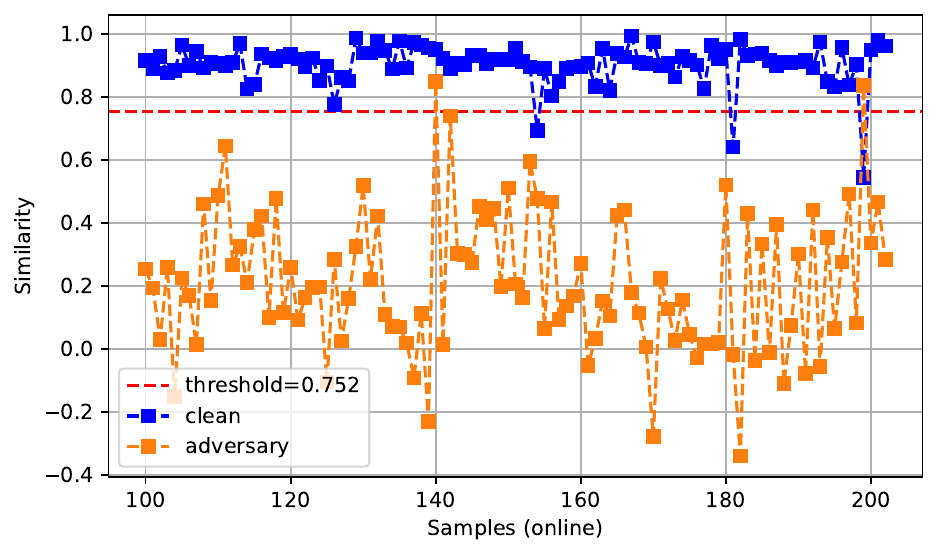}
    \caption{Online: subsample length = 85 (ETTh2)}
    \label{fig_Etth2-85-online234}
\end{subfigure}
\begin{subfigure}[t]{0.32\linewidth}
    \includegraphics[width=\linewidth]{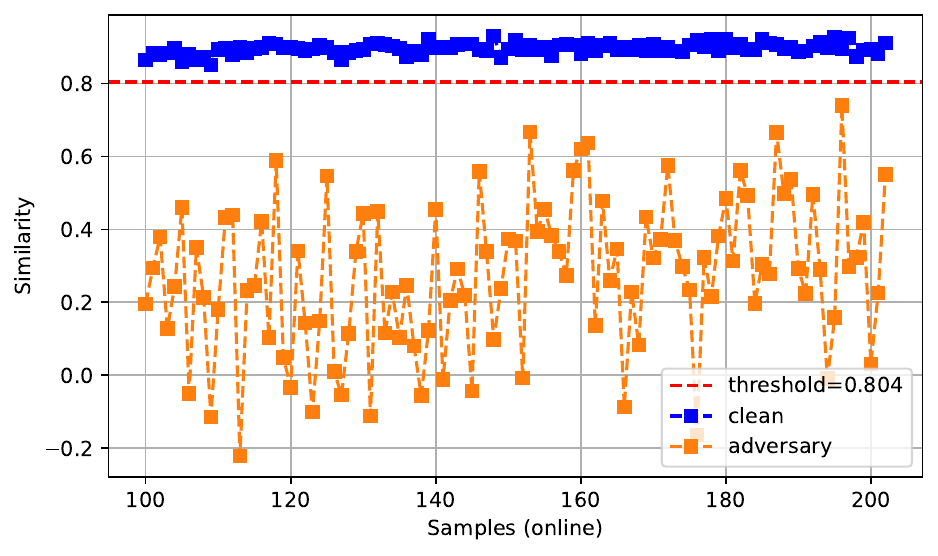}
    \caption{Online: subsample length = 85 (Consum.)}
    \label{fig_Consumption-85-online234}
\end{subfigure}
\begin{subfigure}[t]{0.32\linewidth}
    \includegraphics[width=\linewidth]{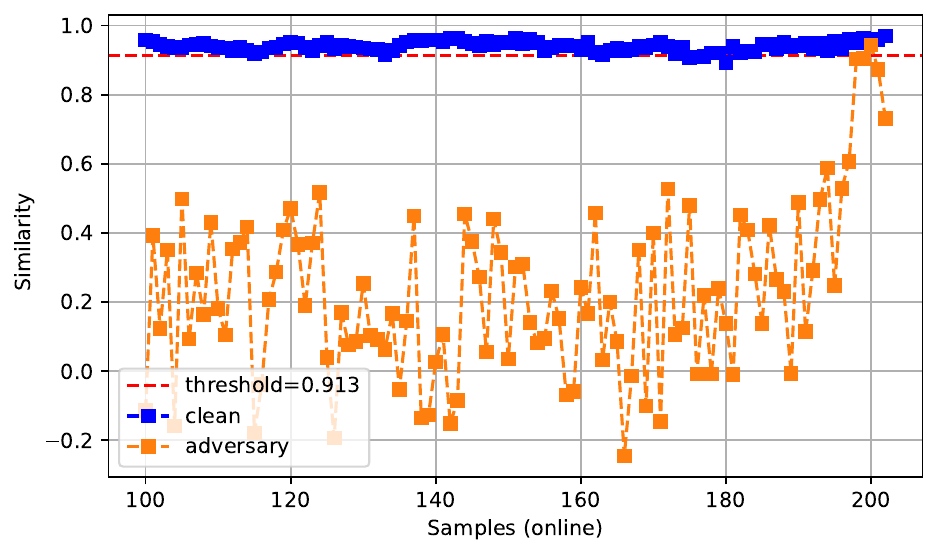}
    \caption{Online: subsample length = 85 (NI)}
    \label{fig_NI-85-online2-3-3-4}
\end{subfigure}

\caption{Offline (top) and online (bottom) detection results for TimesFM across different datasets.}
\label{fig:FM-offline-online}
\end{figure*}

The top row of Figure~\ref{fig:FM-offline-online} shows the offline-determined thresholds, while the bottom row presents the corresponding online detection results for both adversarial and clean samples. The detailed FRR and FAR results are summarized in Table~\ref{tab:FM}. The preset FRR for all datasets is 1.0\%. The thresholds for ETTh2 and Consumption are 0.752 and 0.804, respectively, while the threshold for NI is 0.910.  

For ETTh2, the online FRR for clean samples is 2.9\%, with three samples failing to be correctly detected, and the FAR for adversarial samples is 1.9\%. For the Consumption dataset, both FRR and FAR are 0.0\%, indicating that all clean and adversarial samples are successfully detected. For the NI dataset, the online FRR is 3.9\% and the FAR is 1.0\%.  

\input{Table/Tab_FM}

\subsection{White-Box Adversarial Example Attacks}\label{sec_whichboxAE}
For all previous evaluations, we extensively experimented on black-box adversarial example attacks, which are most realistic in real world. Now we take the white-box AE attacks into account. The white-box attacks require access to the underlying TS-LLM, which is possible when the TS-LLM is a public model or the attacker is an insider. A targeted AE attack aims to force the model to predict a specific (incorrect) value, whereas an untargeted attack seeks to cause the model to output any value different from ground-truth~\cite{papernot2016limitations,carlini2017magnet,gao2024security}. In practical time-series forecasting, the true future values are usually unavailable to the attacker; thus, we assume the adversary cannot access the forecasting ground truth, following~\cite{liu2024adversarial}. Under this assumption, we adopt a targeted attack and set the target value to $0$, optimizing perturbations to reduce forecasting accuracy. 

Three white-box AE attacks of FGSM~\cite{goodfellow2014explaining}, BIM~\cite{kurakin2018adversarial}, and PGD~\cite{madry2018towards} are adopted. It is worth mentioning that none of these AE attacks has been mounted on TS-LLMs so far. TimesFM is used for experiments because it is publicly accessible from Hugging Face, while we use the NI dataset.

As in the black-box setting, a perturbation scale of $0.2$ is adopted to perform white-box AE attacks. The attack performance is shown in Table~\ref{tab:WhiteAttack}, which also includes the DGA black-box attack for comparison. The metrics of MAE, MSE, and $R^2$ are used to evaluate the attack effect. Both MAE and MSE gradually increase for FGSM, BIM, and PGD, while the $R^2$ score decreases significantly. Based on the same setting, the attack effect strengthens from FGSM to PGD, consistent with the expectation that white-box attacks typically exhibit stronger performance than black-box attacks.  

\name is then applied to detect white-box AEs. The FAR and FRR results are detailed in Table~\ref{tab:WhiteDetect}. The subsample length is 85 timesteps, with the original sample length set to 256 timesteps to forecast next 48 timesteps. The offline threshold is 0.913, corresponding to a preset FRR of 1\%. The online FRR is 3.9\%. For adversarial detection, the online FAR is 0\% for both FGSM and BIM, and 1\% for PGD.

\input{Table/tab_WhiteAttack}
\input{Table/tab_WhiteDetect}

\section{Adaptive Attacks}\label{sec_adaptive}

In an adaptive attack, the adversary has full knowledge of the specific defense. However, \name incorporates inherent randomness that complicates such adaptive attacks. This randomness arises from three sources: subsample length, the specific sampling setting (e.g., stride/window), and the particular subsamples used. The attacker cannot control these random choices. 
Note that in the previous section, in some cases, we have intentionally varied subsample length and sampling settings across datasets and TS-LLMs; those experiments show that \name remains largely effective despite such parameter changes, which helps harden against adaptive strategies. Below, we evaluate \name more deliberately by introducing randomness while keeping the \textit{same} NI dataset and the \textit{same} TS-LLM of TimesFM. We then test \name against an adaptive attack, the sparse adversarial attack~\cite{liu2025temporally}, which crafts adversarial examples by perturbing only a fraction of timesteps on the TS-LLM.

\subsection{Sampling Strategy}\label{sec_sampling}
\begin{figure*}[t]
\centering
\begin{subfigure}[t]{0.28\linewidth}
    \includegraphics[width=\linewidth]{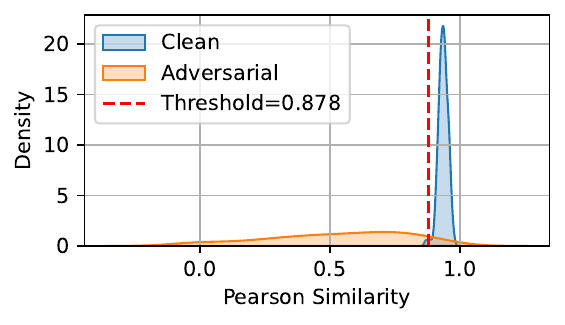}
    \caption{Subsample = 128 timesteps}
    \label{NIFM_128}
\end{subfigure}
\begin{subfigure}[t]{0.28\linewidth}
    \includegraphics[width=\linewidth]{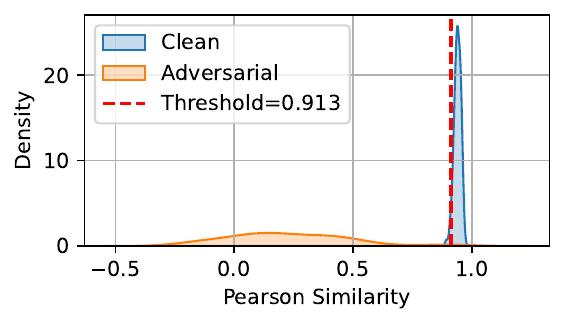}
    \caption{Subsample = 85 timesteps}
    \label{NIFM_85}
\end{subfigure}
\begin{subfigure}[t]{0.28\linewidth}
    \includegraphics[width=\linewidth]{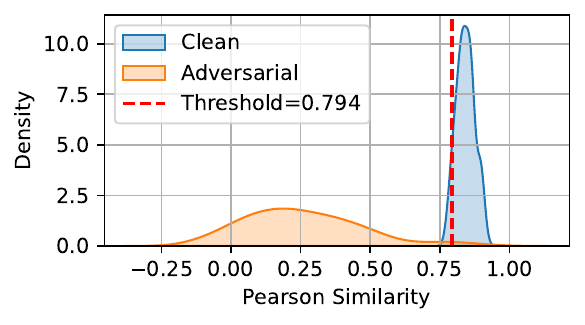}
    \caption{Subsample = 64 timesteps}
    \label{NIFM_64}
\end{subfigure}
\caption{The online similarity distribution across varying subsample
sizes (NI on TimesFM). It shows that there exists a small overlap between two FRR and FAR.}
\label{fig:Dis_subsample}
\end{figure*}

For a given input length, different random sampling strategies and settings can result in varying subsample lengths. Once such randomness is determined by the defender, it remains fixed, similar to setting a random seed, to ensure reproducibility. In Table~\ref{tab:Adsize}, the input sample length is 256 timesteps, while the resulting subsample lengths are 153, 128, 85, and 64. As shown in the table, the offline-determined thresholds are 0.860, 0.878, 0.913, and 0.794 for subsample lengths of 153, 128, 85, and 64, respectively. Figure~\ref{fig:Dis_subsample} then provides a more detailed breakdown of the results for the subsample lengths of 128, 85, and 64.

The defender can randomly choose any of these sampling strategies and apply the corresponding threshold for online detection, as long as it uses the same random seed for sampling. In all cases, the FRR is less than 5.8\% and the FAR is below 6.8\%. This indicates that an adaptive attacker cannot predict which sampling strategy will be adopted for defense.

\input{Table/tab_dis_susample}

\subsection{Used Subsamples}
Even when the subsample length is fixed, the number of subsamples used to determine the threshold can be chosen randomly. In Table~\ref{tab:Ad}, the subsample length is fixed at 180 (from an original input length of 720). Four subsamples of length 180 are generated by taking one sample every three steps, starting from the 1st, 2nd, 3rd, and 4th positions, respectively. The pairwise similarity among these subsamples is then calculated to set the threshold. A single round of comparison among the four subsamples yields $\binom{4}{2} = 6$ pairwise similarity scores.  

The threshold can be determined by averaging any selected subset of these six pairwise similarity scores, allowing flexible aggregation strategies. Table~\ref{tab:Ad} presents an example where different numbers of pairwise similarity scores (from 1 to 6) are used. In all cases, the online detection achieves an FRR of 0.0\% with a maximum FAR of 6.1\%. The total number of possible combinations when selecting at least one pairwise similarity score is: $
\sum_{k=1}^{6} \binom{6}{k} = 2^6 - \binom{6}{0} = 64 - 1 = 63.$

This indicates that an attacker faces significant uncertainty, as the defender can flexibly choose different subsets of pairwise similarity scores to compute thresholds for detection.

\input{Table/tab_Ad_simNum}

\subsection{Sparse Adversarial Attack}\label{sec_sparse} 
In the latest sparse adversarial attack~\cite{liu2025temporally}, perturbations are added only to a fraction of the input timesteps rather than the full sequence. This can be regarded as an adaptive attack to bypass one of \name core insights that adversarial perturbations are optimized over the entire input space.  

In our sparse implementation, 40\% of the input timesteps are perturbed, while the rest timesteps remain intact. For a fair comparison, we ensure that the sparse adversarial attack achieves a similar (though slightly lower) attack effect as the normal attack, where latter all timesteps are perturbed. As shown in Table~\ref{tab:TimesFM_Spaese}, the sparse attack yields an MAE of 0.403, an MSE of 0.311, and an $R^2$ of 61.2\%.  

For \name, the subsample length is set to 64 by uniformly selecting one data point every four timesteps from the original sequence. The offline threshold is determined based on a preset FRR of 1\%. The corresponding online detection results are an FRR of 5.8\% and a FAR of 1.0\%, as reported in Table~\ref{tab:Sparse_detection}.  

In a nutshell, under the same setting and comparable attack effect, the detection performance of \name on sparse adversarial attacks is slightly better (with a 0.9\% decrease in FAR) than on normal attacks. The primary reason is that the perturbation weight per timestep in sparse attacks is higher than in normal attacks; therefore, randomly removing such perturbed timesteps leads to lower similarity among subsamples, making detection easier.

\input{Table/Tab_SparseTimesFM_NI}
\input{Table/Tab_Sparse_detection}

\section{Discussion}
In this section we further discuss several factors that may influence the performance of \name during deployment.

\subsection{FRR vs FAR}\label{sec_frr_vs_far}

FRR refers to the tolerable false rejections of benign samples and is preset in advance. Importantly, there exists a trade-off between FRR and FAR. In security-critical applications, one may sacrifice some user experience in order to suppress FAR, which can be flexibly achieved by adopting a higher preset FRR when determining the detection threshold.

\begin{figure*}[t]
\centering
\begin{subfigure}[t]{0.28\linewidth}
    \includegraphics[width=\linewidth]{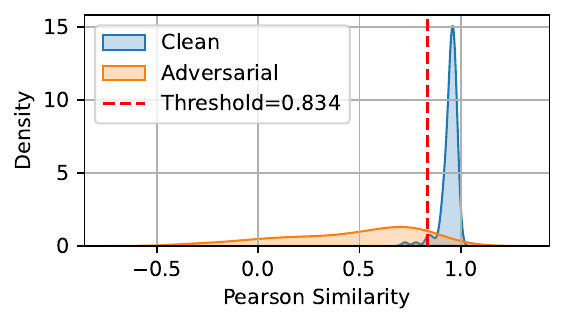}
    \caption{1\% preset FRR}
    \label{Etth2FM0.01}
\end{subfigure}
\begin{subfigure}[t]{0.28\linewidth}
    \includegraphics[width=\linewidth]{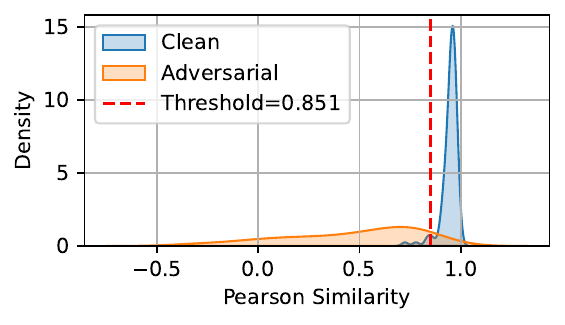}
    \caption{2\% preset FRR}
    \label{Etth2FM0.02}
\end{subfigure}
\begin{subfigure}[t]{0.28\linewidth}
    \includegraphics[width=\linewidth]{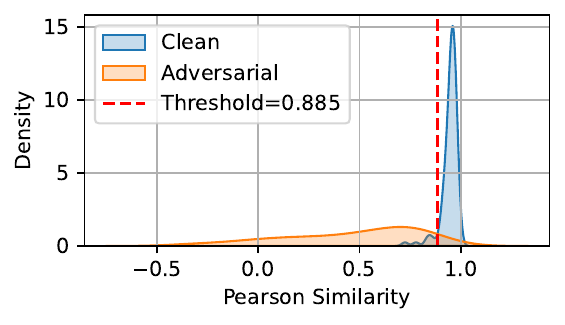}
    \caption{3\% preset FRR}
    \label{Etth2FM0.03}
\end{subfigure}
\caption{The online similarity distribution at different preset FRR tolerances (ETTh2 dataset on TimesFM). It shows that there is a trade-off between FRR and FAR.}
\label{fig:Dis_PresetFRR}
\end{figure*}

To evaluate this trade-off between FRR (usability) and FAR (security), the preset FRR is varied at 1.0\%, 2.0\%, and 3.0\% to determine the threshold and evaluate the online detection results. TimesFM is used for evaluation on the ETTh2 dataset. For each sample with length 256 forecasting the next 48 timesteps, the subsample length is set to 128 timesteps. The detection performance of \name is summarized in Table~\ref{tab:presetFRR} and illustrated in Figure~\ref{fig:Dis_PresetFRR}. When the offline FRR is set to 1.0\%, the online detection achieves an FRR of 1.9\% and an FAR of 8.7\%. Increasing the preset FRR to 2.0\% leads to a slight rise in the online FRR to 3.9\%, while the FAR decreases to 7.8\%. Further raising the preset FRR to 3.0\% results in an online FRR of 5.8\% and a reduced FAR of 2.9\%. These results confirm that tolerating a slightly higher FRR can significantly reduce FAR.

\input{Table/tab_differentFRRETTH2TimesFM}

\subsection{Dynamic Threshold}\label{sec:dynamic}
\begin{figure*}[t]
\centering

\begin{subfigure}[t]{0.28\linewidth}
    \includegraphics[width=\linewidth]{Figure/TimeGPT/Etth2-240-online.pdf}
    \caption{Offline: input length = 240 (ETTh2)}
    \label{fig_Etth2-240-online}
\end{subfigure}
\begin{subfigure}[t]{0.28\linewidth}
    \includegraphics[width=\linewidth]{Figure/TimeGPT/Consumption-360-online.pdf}
    \caption{Offline: input length = 360 (Consum.)}
    \label{fig_Consumption-360-online.pdf}
\end{subfigure}
\begin{subfigure}[t]{0.28\linewidth}
    \includegraphics[width=\linewidth]{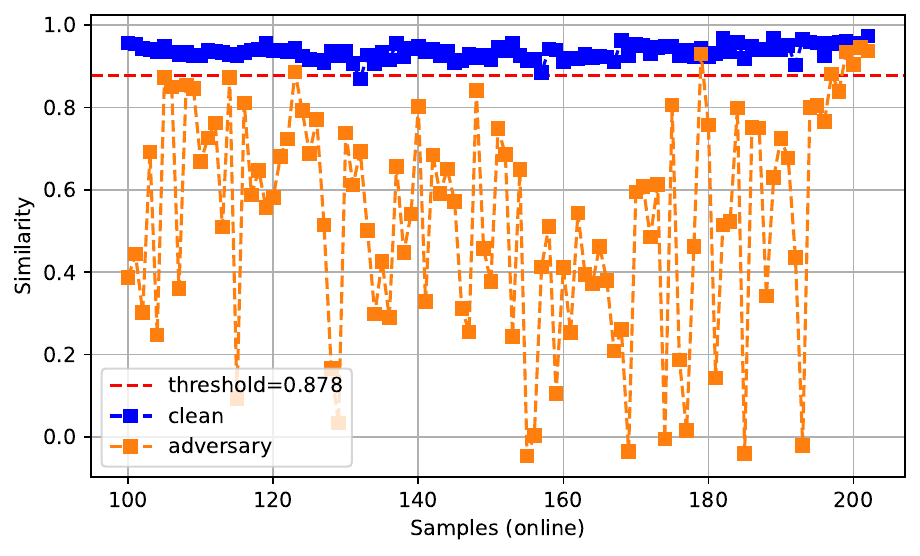}
    \caption{Offline: input length = 128 (NI)}
    \label{fig_NI-128-online}
\end{subfigure}

\vspace{0.3em}

\begin{subfigure}[t]{0.28\linewidth}
    \includegraphics[width=\linewidth]{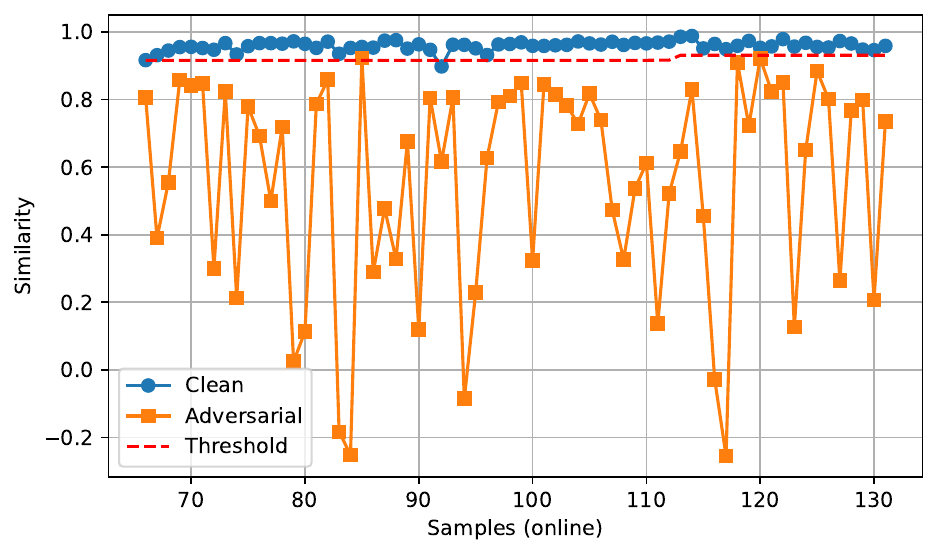}
    \caption{Online: input length = 240 (ETTh2)}
    \label{fig_Etth2-240-online-Dynamic}
\end{subfigure}
\begin{subfigure}[t]{0.28\linewidth}
    \includegraphics[width=\linewidth]{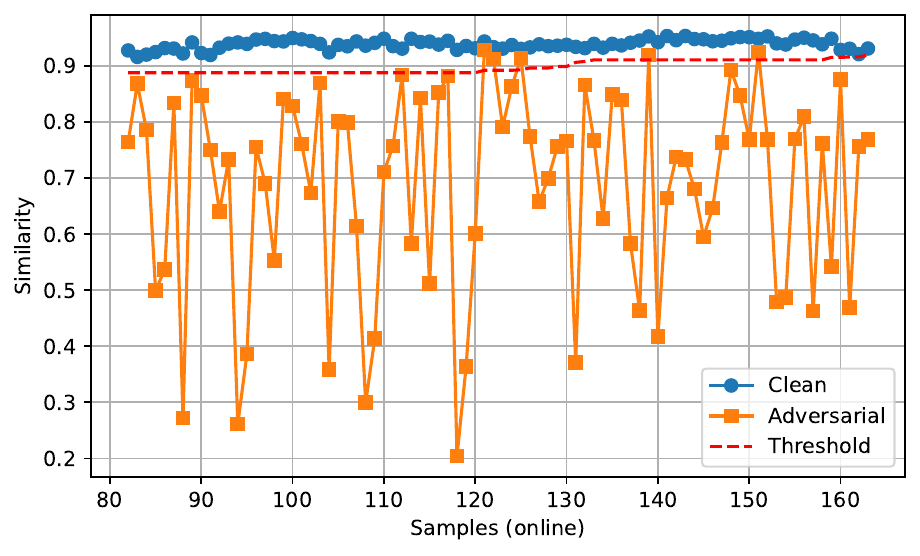}
    \caption{Online: input length = 360 (Consum.)}
    \label{fig_Consumption-360-online-Dynamic-82}
\end{subfigure}
\begin{subfigure}[t]{0.28\linewidth}
    \includegraphics[width=\linewidth]{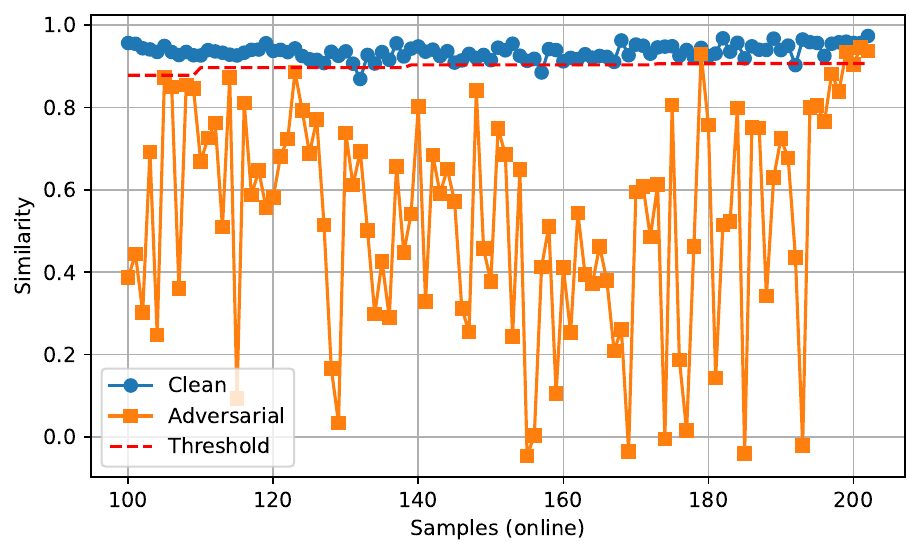}
    \caption{Online: input length = 128 (NI)}
    \label{fig_TimesFM/NI-128-online-Dynamic}
\end{subfigure}

\caption{Online fixed threshold (top) and dynamic threshold (bottom) detection results for across different datasets.}
\label{fig:LLM-Dynamic}
\end{figure*}

When an incoming sample is classified as benign under the online detection, its corresponding similarity can be incorporated to dynamically update the detection threshold. In this update process, the oldest similarity value is discarded to preserve a reference buffer with a constant number of entries (e.g., 62 for ETTh2, 82 for Consumption and 100 for NI) that is consistent with the number of samples employed during the offline phase to determine the fixed threshold. Figure~\ref{fig:LLM-Dynamic} contrasts the dynamic threshold (bottom) with the fixed threshold (top) using the same subsample length across the three test datasets, where ETTh2 and Consumption are evaluated with TimeGPT and NI is evaluated with TimesFM. The details of detection results are shown in Table~\ref{tab:Dynamic}.
Across all datasets, the online detection performance improves significantly when using the dynamic threshold. Specifically, the FAR decreases by approximately 1\% for both the ETTh2 and Consumption datasets compared to the fixed threshold, while the FRR remains unchanged. For the NI dataset, the FAR is reduced by around 2\%, although this has a slight increase of online FRR when using the dynamic threshold instead of the fixed one. In summary, dynamic threshold can always achieve a better balance between online FRR and FAR than that of the fixed threshold, while the fixed threshold is easier to use.

\input{Table/Tab_DynamicThreshold}

\subsection{Similarity Metric}\label{sec:similarity}

\begin{figure*}[t]
\centering
\begin{subfigure}[t]{0.28\linewidth}
    \includegraphics[width=\linewidth]{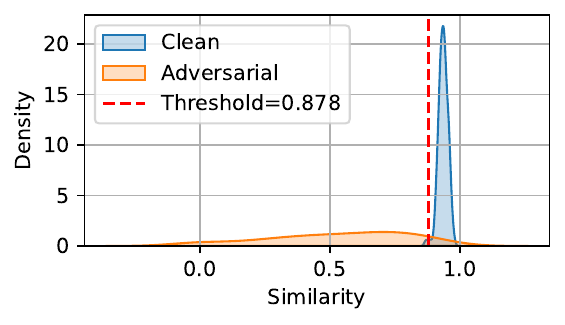}
    \caption{Pearson}
    \label{NIFM128_pearsond}
\end{subfigure}
\begin{subfigure}[t]{0.28\linewidth}
    \includegraphics[width=\linewidth]{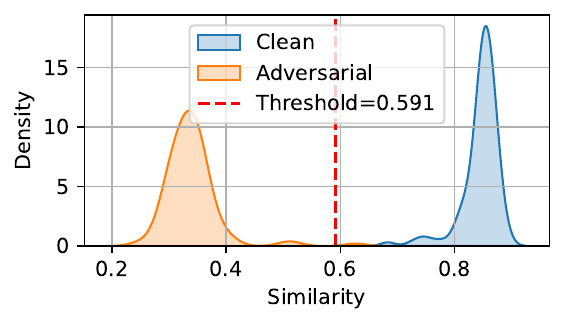}
    \caption{Euclidean}
    \label{NIFM128_euclid}
\end{subfigure}
\begin{subfigure}[t]{0.28\linewidth}
    \includegraphics[width=\linewidth]{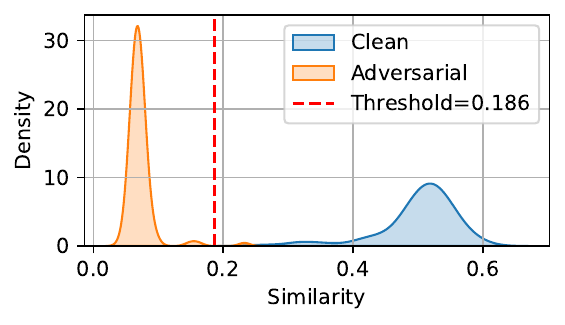}
    \caption{Manhattan}
    \label{Manhattan}
\end{subfigure}
\caption{The online similarity distribution with different similarity metrics (NI on TimesFM).}
\label{fig:dis-metric}
\end{figure*}

In our previous experiments, Pearson similarity was used as the similarity metric. Nonetheless, \name is independent of the choice of similarity metric. 
Here, we evaluate ILID using Euclidean distance~\cite{lines2012shapelet} and Manhattan distance~\cite{wang2013experimental},
each transformed into a similarity measure as similarity = 1/(1 + distance). This simple transformation is to make sure the metric value range is between 0 and 1 for facilitating visualized presentation.
TimesFM is used as the TS-LLM with the NI dataset. Each sample contains 256 timesteps, and the subsample length is set to 128 by selecting either the odd or even elements.
Based on the same offline preset FRR of 1.0\%, the thresholds are 0.591 for the Euclidean similarity metric and 0.186 for the Manhattan similarity metric. 
These thresholds are then applied for detection in the online phase. For both similarity metrics, the online FRR for clean samples is 0\% and the FAR for adversarial samples is 1.0\%, as shown in Figure~\ref{NIFM128_euclid} and Figure~\ref{Manhattan}. These results are not inferior to those obtained using the Pearson similarity metric, where the online FRR was 1.0\% and the FAR was 6.8\%, as shown in Figure~\ref{NIFM128_pearsond}.
Further details are presented in Table~\ref{tab:Similarity Metric}.

\input{Table/tab_Metric}

\subsection{Evaded Adversarial Examples}\label{sec:Undetected}

\begin{figure*}[t]
\centering
\begin{subfigure}[t]{0.28\linewidth}
    \includegraphics[width=\linewidth]{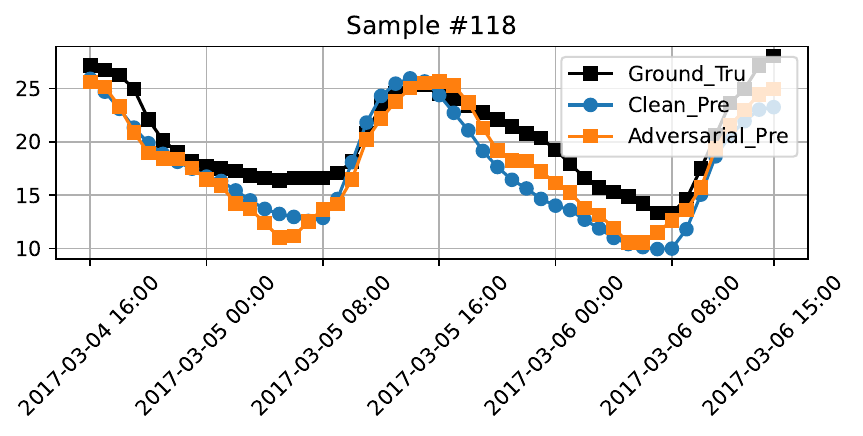}
    \label{Etth2FMSample118-1}
\end{subfigure}
\begin{subfigure}[t]{0.28\linewidth}
    \includegraphics[width=\linewidth]{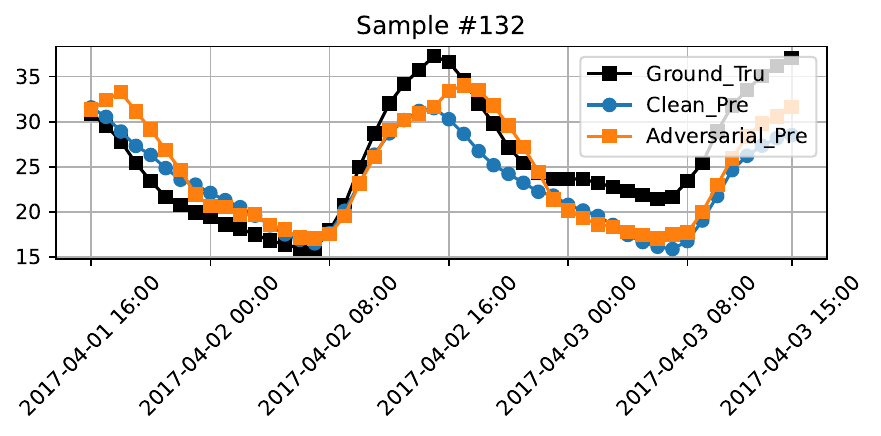}
    \label{Etth2FMSample132-1}
\end{subfigure}
\begin{subfigure}[t]{0.28\linewidth}
    \includegraphics[width=\linewidth]{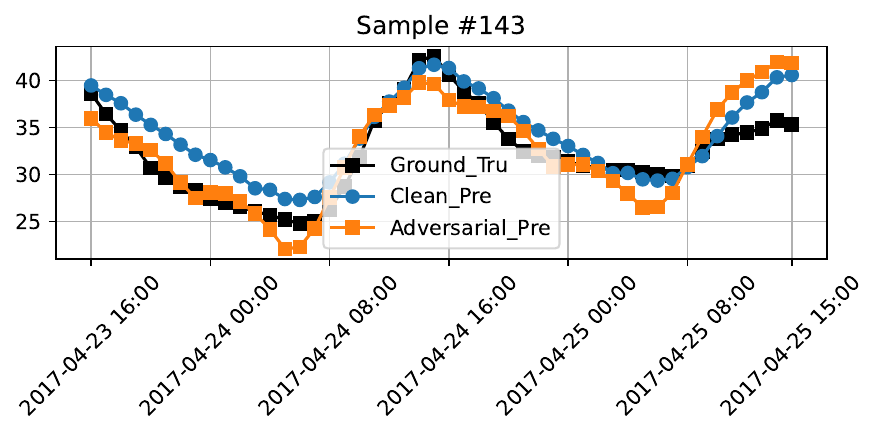}
    \label{Etth2FMSample143-1}
\end{subfigure}
\caption{The forecasting between undetected adversarial sample and the corresponding clean sample  (ETTh2 on TimesFM).}
\label{fig:undteted samples}
\end{figure*}

We further analyze the adversarial examples that evade \name, namely those falsely accepted as benign. 
Our hypothesis is that many of these adversarial inputs may in fact be ineffective, behaving similarly to their clean counterparts. 
To validate this, we compare the forecasting performance of each undetected adversarial sample with its corresponding clean sample. Across datasets, a notable portion of these adversarial examples show no observable attack impact. Table~\ref{tab_evaded_samples} presents the results for the NI and ETTh2 datasets under TimesFM, and the Consumption dataset under TimeLLM. In these tables, blue-highlighted values indicate adversarial samples whose MAE and MSE are equal to or even lower than those of their clean counterparts, providing no evidence of a successful attack. Concretely, for the NI dataset, 2 of 9 samples (\#123 and \#129; 22.2\%) show no attack effect. For the ETTh2 dataset, 3 of 9 samples (\#118, \#132, and \#143; 33.4\%) are similarly ineffective. For the Consumption dataset, 2 of 6 samples (\#100 and \#134; 33.4\%) exhibit no measurable attack. In summary, approximately 30\% of the adversarial examples that bypass \name do so not because they succeed, but because they fail to introduce any meaningful perturbation. This is further illustrated in Figure~\ref{fig:undteted samples}, where three such ETTh2 samples (\#118, \#132, \#143) clearly demonstrate negligible deviation from their clean counterparts.

\input{Table/tab_sample}

\subsection{Broader Application of Forecasting}
Here, we show that \name extends effectively beyond energy-related forecasting tasks. Using the Currency Exchange Rate dataset, which contains daily USD exchange rates for eight countries from 1990 to 2016~\cite{lai2018modeling}, we evaluate \name on a financial forecasting scenario. The experiment uses the USD-to-AUD series, where an adversarial attack is applied to a TimeLLM model forecasting the next 12 steps based on the past 168 timesteps. Table~\ref{tab:exchange} reports the forecasting performance with and without attack. Among the 239 total samples, the first 100 are used to determine an offline threshold of 0.907 with a preset FRR of 1\% and a subsample length of 56. As shown in Table~\ref{tab:ER-detection}, the resulting online FRR and FAR are both 2.9\%. Together, these results highlight the broad applicability and strong generalization ability of \name across diverse domains, including financial time-series forecasting.
\input{Table/Table_exchange_timesLLM}

\input{Table/Tab_exchange_detection}

\section{Conclusion}\label{sec:Conclusion}

This work has proposed \name, which leverages the unique input length independence of TS-LLMs to detect adversarial inputs. Importantly, \name operates without modifying the underlying TS-LLM, functioning instead as a plug-in module that can be seamlessly integrated with TS-LLMs. Extensive experiments on three energy-specific datasets across three TS-LLMs (including both proprietary and open-source models) demonstrate the efficiency and effectiveness of \name in detecting adversarial inputs, under both the latest black-box attacks and well-established white-box attacks.  
Notably, due to its inherent randomness, \name exhibits strong resilience against adaptive attacks; in particular, we demonstrate its effectiveness against a sparse adversarial example attack. Extending and validating \name in other forecasting domains, such as finance, as well as across different modalities, represents a promising direction for future research. This potential lies in the broader fact that foundation models, including LLMs and VLMs, are inherently capable of handling inputs of varying lengths.

\section*{Ethical Considerations}

This work investigates adversarial vulnerabilities in TS-LLMs used for energy forecasting, a domain connected to critical infrastructure. All experiments rely solely on publicly available, non-sensitive datasets, and no operational systems or private information were accessed. While we evaluate both black-box and white-box attacks, we intentionally provide only the technical details required for reproducibility and do not release exploit code or parameters that could enable misuse. 

\section*{Acknowledgment}
This work is supported by CSIRO – National Science Foundation (US) AI Research Collaboration Program (NSF-2302720-CSIRO of RAI4IoE grant).

\bibliographystyle{ACM-Reference-Format}
\bibliography{references}

\end{document}

%% file: _header.tex
\AtBeginDocument{%
  \providecommand\BibTeX{{%
    \normalfont B\kern-0.5em{\scshape i\kern-0.25em b}\kern-0.8em\TeX}}}

\usepackage{amsmath,amssymb,amsfonts,bm,mathtools,mathrsfs}
\usepackage{graphicx,epstopdf,subcaption}
\usepackage{booktabs,multirow,threeparttable,makecell}
\usepackage{algorithm,algpseudocode}
\usepackage{color,xcolor,soul}
\usepackage{xspace,textcomp,gensymb}
\usepackage{stfloats,balance,pifont}
\usepackage[framemethod=TikZ]{mdframed}
\usepackage[title]{appendix}
\usepackage{enumitem}
\usepackage{hyperref}
\hypersetup{colorlinks=true, linkcolor=blue, urlcolor=cyan, citecolor=magenta}
\usepackage{caption}
\newcommand{\name}{\texttt{ILID}\xspace}

\usepackage{amsthm}
\theoremstyle{definition}

\theoremstyle{remark}

\renewcommand\footnotetextcopyrightpermission[1]{}

%% file: Table/Tab_TimeLLM.tex
\begin{table}[t]
\centering
\caption{Detection performance with TimeLLM.}
\label{tab:TimeLLM}
\resizebox{.8\linewidth}{!}{
\begin{tabular}{cccccc}
\toprule
\multirow{2}{*}{Dataset}&
\multirow{2}{*}{Length}
&\multirow{2}{*}{\begin{tabular}{@{}c@{}}Threshold\end{tabular}}
& \multicolumn{1}{c}{Offline} 
& \multicolumn{2}{c}{Online}       \\
\cmidrule(r){4-4}\cmidrule(r){5-6}
& & & FRR  &FRR & FAR  \\
\midrule
ETTh2 &56& 0.852 & 1.4\%& 7.2\% & 8.7\% \\   
Consumption & 240&0.842 & 1.0\% & 1.0\% & 2.0\% \\
NI & 180& 0.908&1.0\%& 3.5\% & 0.9\% \\

\bottomrule
\end{tabular}	
}
\end{table}

%% file: Table/Tab_FM.tex
\begin{table}[t]
\centering
\caption{Detection performance with TimesFM.}
\label{tab:FM}
\resizebox{.8\linewidth}{!}{
\begin{tabular}{cccccc}
\toprule
\multirow{2}{*}{Dataset}&
\multirow{2}{*}{Length}
&\multirow{2}{*}{\begin{tabular}{@{}c@{}}Threshold\end{tabular}}
& \multicolumn{1}{c}{Offline} 
& \multicolumn{2}{c}{Online}       \\
\cmidrule(r){4-4}\cmidrule(r){5-6}
& & & FRR  &FRR & FAR  \\
\midrule
ETTh2 &85& 0.752 & 1.0\%& 2.9\% & 1.9\% \\   
Consumption & 85&0.804 & 1.0\% & 0.0\% & 0.0\% \\
NI & 85& 0.913&1.0\%& 3.9\% & 1.0\% \\

\bottomrule
\end{tabular}	
}
\end{table}

%% file: Table/tab_WhiteAttack.tex
\begin{table}[t]
\centering
\caption{TimesFM forecasting performance w/wo \textit{white-box} adversarial example attack (NI dataset on TimesFM). - means that it is the same as the first row.}
\label{tab:WhiteAttack}
\resizebox{.85\linewidth}{!}{
\begin{tabular}{ccccccc}
\toprule
\multirow{2}{*}{Attack Method}
& \multicolumn{3}{c}{Clean Samples} 
& \multicolumn{3}{c}{Adversarial Samples}       \\
\cmidrule(r){2-4}\cmidrule(r){5-7}
& MAE & MSE & $\text{R}^2$ & MAE & MSE & $\text{R}^2$ \\
\midrule

DGA (Black) & 0.257 & 0.182 & 77.4\% & 0.425& 0.330 & 59.0\% \\ 
FGSM (White)  & - & - & - & 0.534 & 0.454 & 43.5\% \\   
BIM (White) & - & - & - & 0.651 & 0.628 & 21.8\% \\
PGD (White)  & - & - & - & 0.665 & 0.673 & 16.1\% \\
\bottomrule
\end{tabular}	
}
\end{table}

%% file: Table/tab_WhiteDetect.tex
\begin{table}[t]
\centering
\caption{Detection performance with \textit{white-box} adversarial example attack (NI dataset on TimesFM).}
\label{tab:WhiteDetect}
\resizebox{.6\linewidth}{!}{
\begin{tabular}{cccc}
\toprule
\multirow{2}{*}{Attack Method}
& \multicolumn{1}{c}{Offline} 
& \multicolumn{2}{c}{Online}       \\
\cmidrule(r){2-2}\cmidrule(r){3-4}
& FRR  & FRR & FAR  \\
\midrule

DGA (Black) & 1\% & 3.9\% & 1\% \\ 
FGSM (White) & 1\%& 3.9\% & 0\% \\   
BIM (White) & 1\% & 3.9\% & 0\% \\
PGD (White) & 1\% & 3.9\% & 1\% \\
\bottomrule
\end{tabular}	
}
\end{table}

%% file: Table/tab_dis_susample.tex
\begin{table}[t]
\centering
\caption{Detection performance under varying subsample sizes (NI and TimesFM).}
\label{tab:Adsize}
\resizebox{.8\linewidth}{!}{
\begin{tabular}{ccccc}
\toprule
\multirow{2}{*}{Subsample Size}
&\multirow{2}{*}{\begin{tabular}{@{}c@{}}Threshold\end{tabular}}
& \multicolumn{1}{c}{\# of Offline} 
& \multicolumn{2}{c}{\# of Online}       \\
\cmidrule(r){3-3}\cmidrule(r){4-5}
& &  FRR  &FRR & FAR  \\
\midrule
153 & 0.860 & 1.0\%& 1.9\% & 3.9\% \\  
128 & 0.878 & 1.0\%& 1.0\% & 6.8\% \\   
85 & 0.913 & 1.0\% & 3.9\% & 1.0\% \\
64 & 0.794 & 1.0\% & 5.8\% & 1.9\% \\

\bottomrule
\end{tabular}	
}
\end{table}

%% file: Table/tab_Ad_simNum.tex
\begin{table}[t]
\centering
\caption{Threshold and detect results with the different number of pairwise similarity scores (Consumption dataset on TimeGPT).}
\label{tab:Ad}
\resizebox{.75\linewidth}{!}{
\begin{tabular}{ccccc}
\toprule
\multirow{2}{*}{The Number}
&\multirow{2}{*}{\begin{tabular}{@{}c@{}}Threshold\end{tabular}}
& \multicolumn{1}{c}{\# of Offline} 
& \multicolumn{2}{c}{\# of Online}       \\
\cmidrule(r){3-3}\cmidrule(r){4-5}
& &  FRR  &FRR & FAR  \\
\midrule
1 & 0.940 & 1.2\%& 0\% & 1.2\% \\   
2 & 0.912 & 1.2\% & 0\% & 1.2\% \\
3 & 0.843 & 1.2\% & 0\% & 1.2\% \\
4 & 0.836 & 1.2\% & 0\% & 0.0\% \\
5 & 0.775 & 1.2\% & 0\% & 2.4\% \\
6 & 0.692 & 1.2\% & 0\% & 6.1\% \\
\bottomrule
\end{tabular}	
}
\end{table}

%% file: Table/Tab_SparseTimesFM_NI.tex
\begin{table}[t]
\centering
\caption{TimesFM forecasting performance w/wo attack.}
\label{tab:TimesFM_Spaese}
\resizebox{.85\linewidth}{!}{
\begin{tabular}{cccccccc}
\toprule
\multirow{2}{*}{Dataset}
&\multirow{2}{*}{\begin{tabular}{@{}c@{}}Target\end{tabular}}
& \multicolumn{3}{c}{Clean query} 
& \multicolumn{3}{c}{AE query}       \\
\cmidrule(r){3-5}\cmidrule(r){6-8}
& &  MAE & MSE & $\text{R}^2$ &MAE & MSE & $\text{R}^2$  \\
\midrule

NI & NI-MW (normal) & 0.257 & 0.182 & 77.4\% & 0.425 & 0.330 & 59.0\% \\  
NI & NI-MW (sparse)& 0.257 & 0.182 & 77.4\% & 0.403 & 0.311 & 61.2\% \\ 

\bottomrule
\end{tabular}	
}
\end{table}

%% file: Table/Tab_Sparse_detection.tex
\begin{table}[t]
\centering
\caption{Detection performance with sparse attack (NI and TimesFM). }
\label{tab:Sparse_detection}
\resizebox{.75\linewidth}{!}{
\begin{tabular}{ccccc}
\toprule
\multirow{2}{*}{Subsample Size}
&\multirow{2}{*}{\begin{tabular}{@{}c@{}}Threshold\end{tabular}}
& \multicolumn{1}{c}{\# of Offline} 
& \multicolumn{2}{c}{\# of Online}       \\
\cmidrule(r){3-3}\cmidrule(r){4-5}
& &  FRR  &FRR & FAR  \\
\midrule

64 (normal) & 0.794 & 1.0\% & 5.8\% & 1.9\% \\64 (sparse) & 0.794 & 1.0\% & 5.8\% & 1.0\% \\

\bottomrule
\end{tabular}	
}
\end{table}

%% file: Table/tab_differentFRRETTH2TimesFM.tex
\begin{table}[t]
\centering
\caption{Detection performance at different FRR tolerances (Etth2 and TimesFM). }
\label{tab:presetFRR}
\resizebox{.65\linewidth}{!}{
\begin{tabular}{ccccc}
\toprule
\multirow{2}{*}{Subsample Size}
&\multirow{2}{*}{\begin{tabular}{@{}c@{}}Threshold\end{tabular}}
& \multicolumn{1}{c}{\# of Offline} 
& \multicolumn{2}{c}{\# of Online}       \\
\cmidrule(r){3-3}\cmidrule(r){4-5}
& &  FRR  &FRR & FAR  \\
\midrule
128 & 0.834 & 1.0\%& 1.9\% & 8.7\% \\   
- & 0.851 & 2.0\% & 3.9\% & 7.8\% \\
- & 0.885 & 3.0\% & 5.8\% & 2.9\% \\

\bottomrule
\end{tabular}	
}
\end{table}

%% file: Table/Tab_DynamicThreshold.tex
\begin{table}[t]
\centering
\caption{Detection performance w/wo dynamic threshold.}
\label{tab:Dynamic}
\resizebox{.9\linewidth}{!}{
\begin{tabular}{ccccccc}
\toprule
\multirow{2}{*}{Dataset}&
\multirow{2}{*}{Length}
&\multirow{2}{*}{\begin{tabular}{@{}c@{}}Fixed Thr.\end{tabular}}
& \multicolumn{2}{c}{Fixed threshold} 
& \multicolumn{2}{c}{Dynamic threshold}       \\
\cmidrule(r){4-5}\cmidrule(r){6-7}
& & & FRR & FAR  &FRR & FAR  \\
\midrule
ETTh2 &240& 0.915 & 1.5\%& 3.0\% & 1.5\% & 1.5\%\\   
Consumption & 360&0.887 & 0.0\% & 7.3\% & 0.0\%& 6.1\% \\
NI & 128& 0.878&1.0\%& 6.8\% & 2.9\% &3.9\% \\

\bottomrule
\end{tabular}	
}
\end{table}

%% file: Table/tab_Metric.tex
\begin{table}[t]
\centering
\caption{Detection performance with different similarity metrics (NI on TimesFM). }
\label{tab:Similarity Metric}
\resizebox{.6\linewidth}{!}{
\begin{tabular}{ccccc}
\toprule
\multirow{2}{*}{Metric}
&\multirow{2}{*}{\begin{tabular}{@{}c@{}}Threshold\end{tabular}}
& \multicolumn{1}{c}{\# of Offline} 
& \multicolumn{2}{c}{\# of Online}       \\
\cmidrule(r){3-3}\cmidrule(r){4-5}
& &  FRR  &FRR & FAR  \\
\midrule
Pearson & 0.878 & 1.0\%& 1.0\% & 6.8\% \\   
Euclidean & 0.591 & 1.0\% & 0.0\% & 1.0\% \\
Manhattan & 0.186 & 1.0\% & 0.0\% & 1.0\% \\

\bottomrule
\end{tabular}	
}
\end{table}

%% file: Table/tab_sample.tex
\begin{table}[t]
\centering
\caption{Forecasting performance on evaded samples.}
\label{tab_evaded_samples}
\resizebox{.8\linewidth}{!}{
\begin{tabular}{cccccccc}
\toprule
\multirow{2}{*}{Models}
& \multirow{2}{*}{Datasets}
& \multirow{2}{*}{Sample Index} 
& \multicolumn{2}{c}{Clean query} 
& \multicolumn{2}{c}{AE query} \\
\cmidrule(lr){4-5} \cmidrule(lr){6-7}
& & & MAE & MSE & MAE & MSE \\
\midrule
\multirow{6}{*}{TimesFM}
& \multirow{6}{*}{NI}
& 123 & 0.298 & 0.119 & \textcolor{blue}{0.270} & \textcolor{blue}{0.104} \\
& & 197 & 0.486 & 0.311 & 0.693 & 0.588 \\   
& & 199 & 0.917 &1.314 & \textcolor{blue}{0.914} & \textcolor{blue}{1.214} \\
& & 200 & 0.406 & 0.224& 1.245 & 1.773 \\
& & 201 & 0.169 & 0.040 & 0.453 & 0.235 \\
& & 202 & 0.686 & 0.676 & 1.277 & 1.832 \\
\midrule
\multirow{9}{*}{TimesFM}
& \multirow{9}{*}{ETTh2}
& 109 & 0.385 & 0.205 & 0.822 & 0.745 \\
& & 110 & 0.357 & 0.199 & 0.467 & 0.357 \\   
& & 118 & 0.232 & 0.070 &\textcolor{blue}{0.192} & \textcolor{blue}{0.049} \\
& & 132 & 0.314 & 0.134 & \textcolor{blue}{0.280} & \textcolor{blue}{0.101} \\
& & 134 & 0.169 & 0.040 & 0.453 & 0.235 \\
& & 143 & 0.179 & 0.047 & \textcolor{blue}{0.162} & \textcolor{blue}{0.045} \\
& & 150 & 0.328 & 0.171 & 0.420 & 0.309 \\
& & 143 & 0.219 & 0.074 & 0.399 & 0.188 \\
& & 182 & 0.309 & 0.148 & 0.442 & 0.312 \\
\midrule
\multirow{6}{*}{TimeLLM}
& \multirow{6}{*}{Consum.}
& 100 & 0.241 & 0.085 & \textcolor{blue}{0.228} & \textcolor{blue}{0.085} \\   
& & 104 & 0.320 & 0.134 & 0.340 & 0.225 \\
& & 110 & 0.206 & 0.067 & 0.281 & 0.129 \\   
& & 134 & 0.777 & 0.751 & \textcolor{blue}{0.657} & \textcolor{blue}{0.751} \\
& & 142 & 0.604 & 0.399 & 0.659 & 0.587 \\   
& & 180 & 0.275 & 0.108 & 0.454 & 0.299 \\
\bottomrule
\end{tabular}	
}
\end{table}

%% file: Table/Table_exchange_timesLLM.tex
\begin{table}[t]
\centering
\caption{Exchange Rate forecasting performance w/wo attack on TimeLLM.}
\label{tab:exchange}
\resizebox{.9\linewidth}{!}{
\begin{tabular}{cccccccc}
\toprule
\multirow{2}{*}{Dataset}
&\multirow{2}{*}{\begin{tabular}{@{}c@{}}Target\end{tabular}}
& \multicolumn{3}{c}{Clean query} 
& \multicolumn{3}{c}{AE query}       \\
\cmidrule(r){3-5}\cmidrule(r){6-8}
& &  MAE & MSE & $\text{R}^2$ &MAE & MSE & $\text{R}^2$  \\
\midrule

Exchange Rate & Australia & 0.125 & 0.030 & 96.2\% & 0.396& 0.228 & 71.0\% \\ 

\bottomrule
\end{tabular}	
}
\end{table}

%% file: Table/Tab_exchange_detection.tex
\begin{table}[t]
\centering
\caption{Detection performance with Exchange Rate on TimeLLM.}
\label{tab:ER-detection}
\resizebox{.8\linewidth}{!}{
\begin{tabular}{cccccc}
\toprule
\multirow{2}{*}{Dataset}&
\multirow{2}{*}{Length}
&\multirow{2}{*}{\begin{tabular}{@{}c@{}}Threshold\end{tabular}}
& \multicolumn{1}{c}{Offline} 
& \multicolumn{2}{c}{Online}       \\
\cmidrule(r){4-4}\cmidrule(r){5-6}
& & & FRR  &FRR & FAR  \\
\midrule
Exchange Rate &56& 0.907 & 1.0\%& 2.9\% & 2.9\% \\   
\bottomrule
\end{tabular}	
}
\end{table}